\documentclass[aps,pre,superscriptaddress,floatfix,amsmath,amssymb,longbibliography,11pt]{revtex4-1}

\usepackage{xcolor}
\usepackage{amssymb}
\usepackage{amsmath}
\usepackage{bm}
\usepackage{url}
\usepackage{mathtools}
\usepackage{soul}
\usepackage{graphicx}
\usepackage{nicefrac}

\DeclareMathAlphabet\mathbfcal{OMS}{cmsy}{b}{n}

\begin{document}

\title{Reinforcement learning for pursuit and evasion of microswimmers at low Reynolds number}\thanks{Version accepted for publication (postprint) on Phys. Rev. Fluids 7, 023103 (2022) – Published 23 February 2022 DOI:10.1103/PhysRevFluids.7.023103}

\author{Francesco Borra}
\thanks{Present address: Laboratory of Physics of the Ecole Normale Sup\'erieure, CNRS UMR8023, 24 rue Lhomond, 75005 Paris, France}
\affiliation{Dipartimento di Fisica, Universit\`a ``Sapienza''  Piazzale A. Moro 5, I-00185 Rome, Italy}

\author{Luca Biferale} \affiliation{Department of Physics and INFN, University of Rome Tor Vergata, 00133 Rome, Italy}

\author{Massimo Cencini}
\thanks{Corresponding author}
\email{massimo.cencini@cnr.it}
\affiliation{Istituto dei Sistemi Complessi, CNR,  00185 Rome, Italy and INFN ``Tor  Vergata''}

\author{Antonio Celani}
\thanks{Corresponding author}
\email{celani@ictp.it}
\affiliation{Quantitative Life Sciences, The Abdus Salam International 
Centre for Theoretical Physics - ICTP, Trieste, 34151, Italy}

\begin{abstract}
  We consider a model of two competing microswimming agents engaged in
  a pursue-evasion task within a low-Reynolds-number
  environment. Agents can only perform simple maneuvers and sense
  hydrodynamic disturbances, which provide ambiguous (partial)
  information about the opponent's position and motion. We frame the
  problem as a zero-sum game: The pursuer has to capture the evader in
  the shortest time, while the evader aims at deferring capture as
  long as possible.  We show that the agents, trained via adversarial
  reinforcement learning, are able to overcome partial observability
  by discovering increasingly complex sequences of moves and
  countermoves that outperform known heuristic strategies and exploit
  the hydrodynamic environment.
\end{abstract}


\maketitle
\section{Introduction}
Aquatic organisms can detect moving objects by sensing the induced
hydrodynamic disturbances
\cite{triantafyllou2016biomimetic,takagi2018directional,tuttle2019going}. Such
an ability is crucial in prey-predator interactions and for
navigation, especially in murky or dark waters, as for the blind
Mexican cavefish \cite{lloyd2018evolutionary}. Fishes have developed
the lateral line, a mechanosensory system very sensitive to water
motions and pressure gradients
\cite{montgomery1997lateral,bleckmann2009lateral,kanter2003rheotaxis}.
Planktonic microorganisms, inhabiting a low-Reynolds-number
environment, have antennae and setae to sense hydrodynamic signals
produced by predators and preys
\cite{kiorboe1999predator,doall2002mapping}.

Abstracting away from specific mechanisms developed by aquatic
organisms, the problem of pursue-evasion in microswimmers guided by
hydrodynamic cues poses substantial difficulties rooted in the physics
of the ambient medium.  At low Reynolds numbers, flow disturbances are
generally weak and rich of symmetries \cite{happel2012low} leading to
ambiguities about the signal source location especially if distant
from the receiver
\cite{kiorboe1999predator,takagi2018directional,tuttle2019going}.
Moreover, hydrodynamics has dynamical effects, as the disturbances
generated by one microswimmer alter the other
motion. Consequently, an agent's strategy inevitably
  affects the opponent dynamics and strategies. It is thus crucial to
  understand how agents' strategies co-evolve by competing against one
  another \cite{hein2020algorithmic}, which necessitates going beyond just
  escaping from a prescribed pursuit strategy or pursuing a
  nonresponsive moving target
  \cite{domenici2011animal,nahin2012chases}.  Which pursuit-evasion
  strategies can be devised in such dynamic, partially observable
  environments?  How do they coevolve while competing?  Can
  hydrodynamics be exploited and how?  How do they compare with
  strategies based on visual cues?

Here, we formulate the problem of prey-predator microswimmers
in a game-theoretic framework \cite{hofbauer1998evolutionary}, a
natural setting to model the emergence of adversarial strategies
\cite{hein2020algorithmic}.  As we are interested in the learning and
evolution of strategies and not in fine tuning on specific details of
the two microswimmers, we choose a simplified hydrodynamics.  Inspired
by recent applications of multi agent reinforcement Learning (MARL)
\cite{sutton2018reinforcement} to hide-and-seek contests
\cite{baker2019emergent,chen2020visual}, we explore its use as a
general model-free framework for discovering effective
chase-and-escape strategies at low Reynolds number. Reinforcement
learning (RL) approaches rely on trial and error to improve the
quality of the decisions made by an agent -- here a microswimmer --
and has been already applied in numerical and experimental study of
navigation in complex fluid environments
\cite{biferale2019zermelo,alageshan2020machine,reddy2016learning,colabrese2017flow,verma2018efficient,mirzakhanloo2020active,cichos2020machine,qiu2021active,reddy2018glider,muinos2021reinforcement}. We
show that RL is able to discover complex strategies, evolving during
the different phases of the adversarial learning, and thus depending
on the combined training history. The discovered strategies
efficiently overcome the limitations imposed by the partial
observability. In particular, pursuer strategies are shown to
outperform a heuristic baseline policy. \textcolor{black}{Moreover, we show that
  the main strategies discovered by RL are explainable and for some of them we provide an analytical description, which allows us to rationalize how the pursuer
  overcomes the difficulties due to partial information.}

\textcolor{black}{The material is organized as follows. In
  Sec.~\ref{sec:model} we present the model. In
  Sec.~\ref{sec:learning} we discuss the basic ideas of reinforcement
  learning applied to our model and some detail on the
  implementation. In Sec.~\ref{sec:results} we present the results,
  while Sec.~\ref{sec:conclusions} is devoted to discussions and
  conclusions. Some more technical material is presented in the
  Appendices and details on the numerical implementation of the
  reinforcement learning algorithm are discussed in supplementary
  material \footnote{See {S}upplemental {M}aterial [url] for
    details on the implemented Reinforcement
    Learning algorithm including a pseudo-code, for further exploration with different parameters and with rotational noise inlcuded,and for the captions
    of Supplementary movies.}.}

\section{Model \label{sec:model}}

\subsection{Game theoretic formulation \label{sec:modelGame}}
\textcolor{black}{The basic settings of the game-theoretic formulation
  of the problem are shown in Figs.~\ref{fig:1}(a) and 1(b). Agents have a
  limited maneuverability and partial information on the opponent via
  hydrodynamic cues, which we choose to be the gradients of the
  velocity field [Fig.~\ref{fig:1}(a)].  The two swimming agents play
  the following zero-sum game [Fig.~\ref{fig:1}(b)]: They start at
  distance $R_0$ with random heading directions. At each decision time
  $\tau$ each agent senses the hydrodynamic field and chooses an
  action (steer left/right or go straight).}  The pursuer ($p$) aims
at reaching the capture distance $R_c$ from the evader ($e$) in the
shortest possible time, while the latter has to keep the pursuer at
bay (at distance $R >R_c$). The game terminates either on capture
(pursuer wins) or if its duration exceeds a given time $T_{max}$
(evader wins).  \textcolor{black}{While playing many games the agents
  are trained via via reinforcement learning (see
  Sec.~\ref{sec:learning})}

\subsection{Modeling the agents\label{sec:modelSwimmers}}
\begin{figure}[t!]
\centering
\includegraphics[width=0.9\columnwidth]{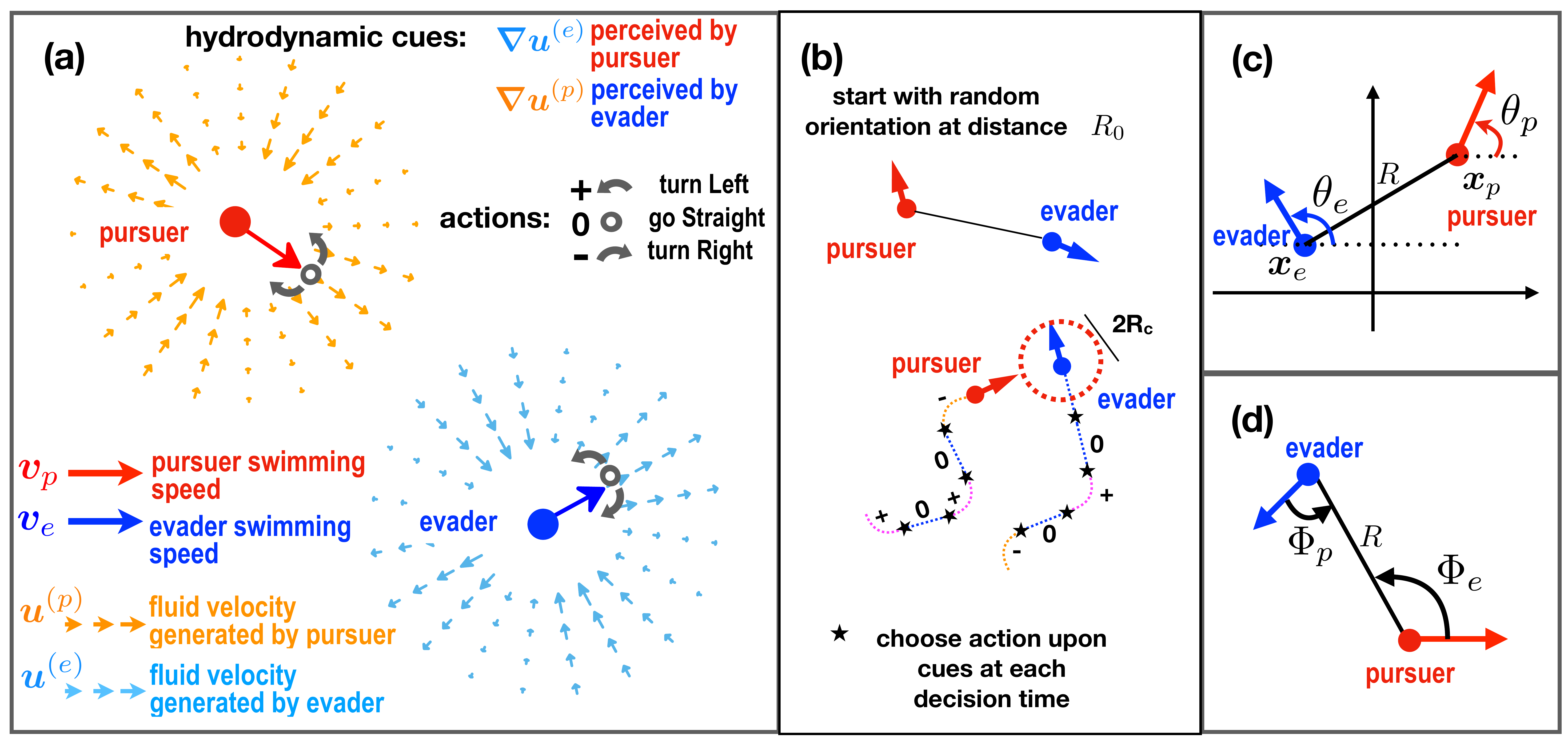}
\caption{\textcolor{black}{(Color online) Model illustration. (a) Basic
    elements: The pursuer ($p$, red)/evader ($e$, blue) swims with
    speed $v_{p/e}$ generating a velocity field $\bm u^{(p/e)}$, which
    drags the other agent and offers a cue to the other agent on the
    relative position and orientation via its gradients, $\bm \nabla
    \bm u^{(p/e)}$. Agents only have a limited control on their
    heading directions -- the actions. (b) Sketch of a game: The game
    starts with the agents at distance $R_0$ and the pursuer/evader goal is
    to min/maximize the time their distance reaches the capture value
    $R_c$ within a given time horizon.  Agents move in the plane,
    every $\tau$ time-unit they choose to maintain or turn left/right
    their heading direction on the basis of the cues they receive.
    (c) Geometry of the problem in a fixed frame of reference with
    indicated the heading angles. (d) Bearing angle $\Phi_{e/p}$
    corresponding to the angular position of an agent with respect to
    the heading direction of its opponent.}
  \label{fig:1}}
\end{figure}

For simplicity, we model the agents as ``pusher'' discoids
\textcolor{black}{in an idealized two-dimensional environment disregarding any effect due to walls or confinements and} in the absence
of external flows. By swimming, they generate a velocity field modeled
as a force dipole moving with speed $v_\alpha$ with $\alpha\!=\!e,p$
[Fig.~\ref{fig:1}(a)]. The force-dipole approximates well the far field
of many microorganisms \cite{lauga2009hydrodynamics}.  Near-field
corrections, depending on details, are not implemented as we are not
interested in tuning the model to a specific swimming mechanism,
though they can matter in close encounters
\cite{ishimoto2020regularized}.  Besides self-propulsion each
microswimmer is advected and reoriented by the flow generated by the
other.  Every $\tau$ time units, i.e., at each decision time, agents can
steer by imparting a torque, resulting in an angular velocity
$\Omega_\alpha$.  Thus the position $\bm x_\alpha$ and heading
direction $\bm n_\alpha \!=\!(\cos\theta_\alpha,\sin\theta_\alpha)$
evolve as
\begin{eqnarray}
  \dot{\bm x}_\alpha &=& v_\alpha \bm n_\alpha +\bm u^{(\beta)}\label{eq:x} \\
  \dot{\theta}_\alpha &=& \Omega_\alpha +\omega^{(\beta)}/2 \label{eq:th}\,,
\end{eqnarray}
\textcolor{black}{ where $\bm u^{(\beta)}(\bm x)$ and $\omega^{(\beta)}(\bm x)(=\bm
\nabla \times \bm u^{(\beta)}(\bm x))$ are the velocity and vorticity
field at position  at position $\bm x_\alpha$, generated by the
opponent agent $\beta$ in $\bm x_\beta$ with heading orientation
$\theta_\beta$, see Fig.~\ref{fig:1}c.}

\textcolor{black}{As detailed in  Appendix~\ref{app:stokes}, we can write $\bm
u^{(\beta)}(\bm x)\!=\!(\partial_y,-\partial_x)\Psi(\bm x\!-\!\bm
x_\beta;\theta_\beta)$, with the stream function
$\Psi=\nicefrac{D_\beta}{2} \sin(2\phi-2\theta_\beta)$, where $\bm
x-\bm x_\beta=|\bm x-\bm x_\beta| (\cos\phi,\sin\phi)$ and $D_\beta$
denotes the dipole intensity of agent $\beta$. We consider
$D_\beta>0$, i.e., pusher like microswimmers
\cite{lauga2009hydrodynamics}. The velocity in Eq.~(\ref{eq:x}) is
then obtained by deriving the stream function in $\bm x=\bm x_\alpha$,
corresponding to $\phi=\phi_\alpha$ [see Fig.~\ref{fig:geom}(a) for the
notation on the angles with respect to a fixed frame of
reference]. While the vorticity in Eq.~(\ref{eq:th}) is 
$\omega^{(\beta)}=2D_\beta/R^2\sin(2\phi_\alpha-2\theta_\beta)=
2D_\beta/R^2\sin(2\phi_\beta-2\theta_\beta)$, with $R=|\bm
x_\alpha-\bm x_\beta|$ and the second equality stemming from
$\phi_\beta=\phi_\alpha+\pi$ [Fig.~\ref{fig:geom}(a)].}

\subsection{Modeling the hydrodynamic cues\label{sec:modelCues}}
\textcolor{black}{As already discussed,} we assume an agent can only
sense the gradients of the velocity field generated by its opponent,
similarly to what copepods do with sensory setae
\cite{kiorboe1999predator}.  \textcolor{black}{Since agents have no
  notion of an external frame of reference we assume that they
  perceive the gradients in their own frame of reference, i.e.,
  projected along the angents' swimming direction. In this frame of
  reference the three independent components (vorticity and longitudinal
  and shear strain) of the velocity gradients read
\begin{eqnarray}
  \omega^{(\beta)}\! &=& \partial_x u^{(\beta)}_y\!-\!\partial_y u^{(\beta)}_x=\frac{2D_\beta}{R^2} \sin (2\Phi_\beta-2\Theta_\beta) \label{eq:vort}\\
  \mathcal{L}^{(\beta)}\! &=& \partial_x u^{(\beta)}_x\!=\!-\partial_y u^{(\beta)}_y=-\frac{D_\beta}{R^2} \cos (4\Phi_\beta\!-\!2\Theta_\beta) \label{eq:long}\\
\mathcal{S}^{(\beta)}\! &=& \frac{1}{2}(\partial_x u^{(\beta)}_y\!+\!\partial_y u^{(\beta)}_x)\!=\!-\frac{D_\beta}{R^2} \sin(4\Phi_\beta\!-\!2\Theta_\beta)\,. \label{eq:shear}
\end{eqnarray}
They depend on agents' distance ($R$), relative heading
$\Theta_\beta\!=\!\theta_\beta-\theta_\alpha$ [see
Fig.~\ref{fig:geom}(b)] and angular position of
$\beta$ with respect the heading direction of $\alpha$,
$\Phi_\beta\!=\!\phi_\beta-\theta_\alpha$, i.e., the bearing angle
[Fig.~\ref{fig:1}(d)], as called in the pursuit-evasion-games
language~\cite{nahin2012chases}.}

\textcolor{black}{In general, gradients are symmetric with respect
to parity, i.e., to the combined transformation $\Theta_\beta\!\to\!
\Theta_\beta+\pi$ and $\Phi_\beta \!\to\!  \Phi_\beta+\pi$. The force dipole
case is even more degenerate as, owing to the fore-aft symmetry,
either of the two transformations leaves the gradients unchanged, due to the nematic nature of dipoles. Such
symmetries result in ambiguities in the identification of the position
and orientation of the opponent, akin to the $180^o$ ambiguity in fish
hearing \cite{wubbels1997neuronal}.  Memory of past detections and/or
multiple hydrodynamical cues can, in principle, mitigate such ambiguities which,
however, typically persist at large distances
\cite{sichert2009hydrodynamic,triantafyllou2016biomimetic,takagi2018directional}. In
spite of its simplicity the model is thus rich enough to represent the
typical observability limitation inherent to  organisms that can
  only perceive the gradients of the velocity field.}

\section{Learning to pursue and evade through reinforcement \label{sec:learning}}
To set up a learning framework, we need to identify: a set of
observations, $o$, that an agent perceives and uses to infer the
opponent's state; the actions, $a$, through which it can implement its
strategy; and the rewards, $r$, to evaluate its actions.  The learning
task here is to find an optimal reactive policy, $\pi^*(a|o)$, that
associates actions to observations in order to maximize the expected
cumulative rewards. In our setting, the environmental state (relative
position and heading) is only partially observable through the
velocity gradients\cite{jaakkola1995reinforcement}. The actions $a \in
{\cal A}\!=\!\{0,+,-\}$ [Fig.~\ref{fig:1}(a)] correspond to the three
angular velocities $\Omega_\alpha\!=\!0,+\varpi_\alpha,-\varpi_\alpha$
agent $\alpha$ can choose to control its orientation. Once actions are
taken, the agents evolve for a time $\tau$ with the dynamics
(\ref{eq:x}) and (\ref{eq:th}) and a reward is issued. In this zero-sum
game, the currency is the elapsing time: The pursuer/evader receives a
reward $r\!=\!\mp 1$ at the end of each decision time. After each
action, the agents update their policy by combining past and new
information with the issued reward. In the new state, gradients are
sensed again, new actions are taken and rewards received; the cycle
repeats itself until the terminal state is achieved, with either the
pursuer (if $R\leq R_c$) or the evader winning (if the game duration exceeds $T_{max}$). \textcolor{black}{The total return accumulated by the pursuer/evader in an episode is therefore $\mp T$ where $T$ is the duration of the episode itself}.

\subsection{Reinforcement Learning algorithm\label{sec:learningAlgo}}
Among the many approaches to MARL we adopt a natural actor-critic
architecture (see \cite{bhatnagar2009natural,grondman2012survey} and
supplementary material \cite{Note1} for details) because of its theoretical guarantees and
connection with evolutionary game theory
\cite{hennes2019neural,hofbauer1998evolutionary}. In this class of
algorithms, locally optimal solutions are sought by means of
stochastic gradient ascent in policy space. Natural gradients are
used, by virtue of their covariance with respect to the metric defined
by the Fisher information \cite{amari1998natural}.  Real organisms
process the environmental cues with their nervous system that encodes
the policy, e.g., in fishes dedicated neurons control escape responses
\cite{eaton2001mauthner}. Such neural encoding can be emulated by
artificial neural networks \cite{arulkumaran2017deep}.  Here, in the
interest of explainability, we opted for an explicit parametrization
of the policy in terms of few selected features of the observations.
Dropping the agent indices for simplicity, we set
$$\pi(a|o)=\frac{\exp(\mathbfcal{F}(o)\cdot \bm \xi_a)}{\sum_{a'}
  \exp({\mathbfcal{F}(o)\cdot \bm \xi_{a'}})}\,,$$ where $a,a'\in
\mathcal{A}$, $\mathbfcal{F}(o)$ are features that encode the
observations $o$, and $\bm \xi_a$ the learning parameters.

By combining the velocity-gradient components, we chose to extract the
following observables ($o$) (see Appendix~\ref{app:features}): the
vorticity $\omega$; a proxy for the agents' distance, $\hat{R}\propto
1/R^2$; and a linear combination of heading and bearing angle
$\gamma=4\Phi\!-\!2\Theta$.  As features, $\mathbfcal{F}(o)$, we used
the raw observables $\omega$ and $\hat{R}$, and the first and second
harmonics of angle $\gamma$. To encode for the heading direction, we
include some short-term memory by combining a few past observations.
As discussed in Appendix~\ref{app:features}, exploratory studies with more
features did not give qualitatively different results from the minimal
setting described above. Moreover, eliminating memory yields the same
strategies, \textcolor{black}{which indicates that
 a more sophisticated exploitation of memory is needed.}

\subsection{Training scheme\label{sec:learningScheme}}
To better interpret the evolution of strategies and
  counter-strategies, we organized learning in phases (each made of
  $M\!=\!5\times 10^3$ episodes) where agents alternately improve their
  policies. Assuming no prior knowledge, agents start their training
with a random policy, $\pi(a|o)\!=\!1/|\mathcal{A}|\!=\!1/3$ for all
$o$.  At first, the pursuer learns with the evader's policy frozen,
and then the evader learns against the pursuer policy from the previous
phase, and so on.  Episodes start with agents at a distance
$R_0\!=\!1$ and random heading directions, and end either on capture
($R\!\leq\!  R_c\!=\!0.05 R_0$) or when time exceeds the cap
$T_{max}\!=\!50T_0$, where $T_0\!=\!R_0/v_e$ is estimated in terms of
the evader speed and initial distance.  We fixed the evader speed at
$v_e\!=\!0.1$ and angular velocity $\varpi_e\!=\!3$. For the pursuer,
we chose $(v_p,\varpi_p)\!=\!(0.15,4.5)$ which gives a slight speed
advantage maintaining the same steering ability (same curvature radius
$v_p/\varpi_p\!=\!v_e/\varpi_e$). The intensity of the force dipole is taken to be equal for both agents $D_p=D_e=0.03$. With this choice hydrodynamic
velocity dominates over swimming at distances $R\lesssim R_0$. The
decision time is $\tau\!=\!0.01 T_0$ for both agents.

\section{Results\label{sec:results}}

\begin{figure*}[t!]
\centering
\includegraphics[width=1\textwidth]{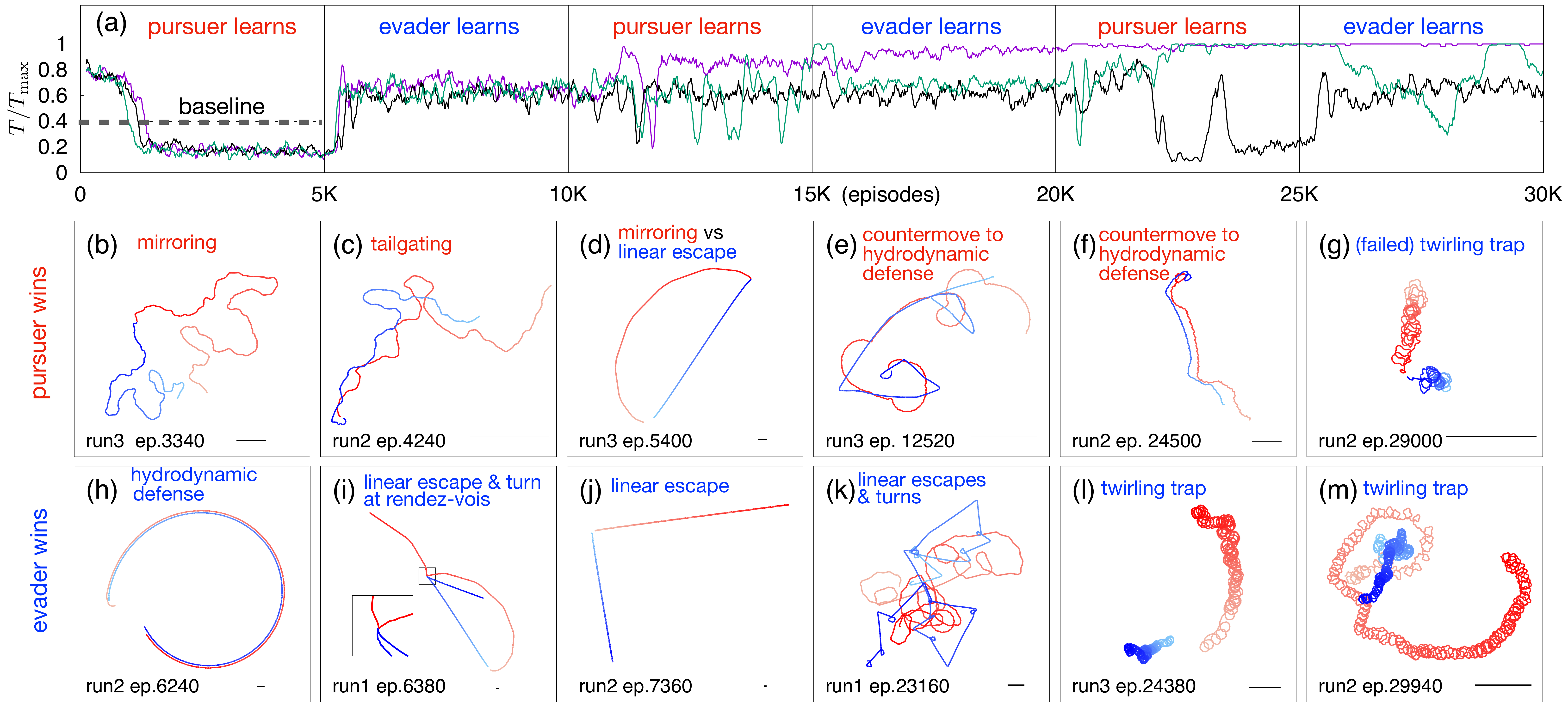}
\vspace{-0.4truecm}
\caption{(Color online) Co-evolving strategies in the first six
    training cycles.  (a) Running average (over 100 episodes) of
    normalized episode duration $T/T_{max}$ for three realizations of
    learning: run1, run2 and 3 (purple, green and black curves). The
    dashed horizontal gray line represents a heuristic baseline value
    as discussed in Sec.~\ref{sec:baseline}. [(b)-(g)] Winning pursuit strategies: (b)
    mirroring, (c) tailgating, (d) mirroring vs linear escape with a
    rendez-vous, [(e) and (f)] tailgating with countermoves to hydrodynamic
    defense, and (g) failing twirling on mirroring. [(h)-(m)] Winning evasion
    strategies: (h) hydrodynamic defense, (i) linear escape with turn
    and hydrodynamic collision at rendez-vous, (j) linear escape
    against mirroring, (k) linear escapes and turns inducing pursuer
    switches between mirroring and tailgating at distance, and [(l) and (m)]
    twirling trap.  Red/blue denotes pursuer/evader trajectories, time
    runs from lighter to darker color (apparent close encounters
    actually take place at different times); run/episode labeled on
    each panel; the bottom-right bar displays the unit length.
  \label{fig:2}}
\end{figure*}

Our main results are summarized in Fig.~\ref{fig:2}: Figure~2(a) shows
the \textcolor{black}{running average of} normalized game duration
$T/T_{max}$ \footnote{Note that each point represents the average over
  the previous and following 50 episodes, so it is not immediate to
  recognize those episodes in which the evader wins, i.e. in which
  $T/T_{max}=1$.} in the first six learning phases for three
independent learning experiments; Figs.~2(b)-2(g) and Figs.~2(h)-2(m)
display some representative examples of pursuer and evader winning
strategies, respectively.  Cycles 1 and 2 are quite reproducible: The
pursuer discovers ways to rapidly catch the evader which, in turn,
finds ways to counteract.  Conversely, cycles 3-6 are characterized by
a higher variability: Agents seem to acquire and lose good policies
also within their own learning turn, and we see cases (run1 in
Fig.~\ref{fig:2}a) in which the evader eventually dominates the
game. We hypothesize that such variability arises from a combination
of insufficient hyperparameters tuning \footnote{We use a fixed
  learning rate instead of an adaptive one, and possibly, due to the
  need to explore, a larger number of episodes per turn would be
  necessary.}  and/or subtle instabilities in the learning
algorithm. Notwithstanding these limitations, many aspects of the
learned strategies are reproducible and, to some extent, physically
explainable as discussed below.  \textcolor{black}{The effect of a
  variation of the parameters and the addition of rotational noise is
  discussed in Sec. II of supplementary material \cite{Note1}.}

\subsection{Pursuit strategies: Mirroring and tailgating\label{sec:mt}}
\begin{figure}[t!]
\centering
\includegraphics[width=0.5\textwidth]{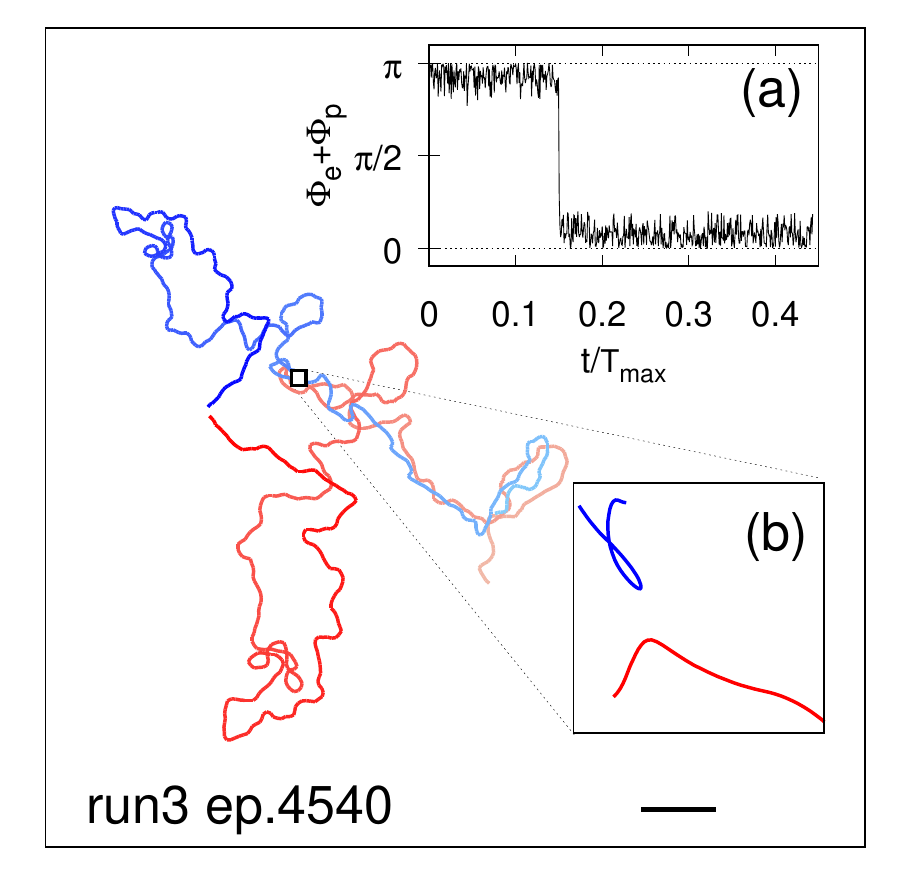}
\vspace{-0.5truecm}  
\caption {(Color online) Switching between tailgating to
  mirroring. Inset (a): Sum of bearing angles $\Phi_e\!+\!\Phi_p$ vs
  normalized time.  Proximity and evader turning [inset (b)] triggers
  the switch $\Phi_e\!+\!\Phi_p \!\approx \!\pi \!\to \! 0$
  (tailgating $\to$ mirroring) at $t/T_{max}\!\approx\! 0.15$. 
  \label{fig:3}}
\end{figure}

In its first learning phase, the evader executes a random
cue-insensitive policy, while the predator learns to pursue its prey
either ``mirroring'' its actions [Fig.~\ref{fig:2}(b)] or ``tailgating'' it
[Fig.~\ref{fig:2}(c)]. When the pursuer approaches the evader, a switch
between the two strategies can sometimes be observed presumably due to
hydrodynamical effects overcoming self-swimming at these distances
combined to evader turning [Fig.~\ref{fig:3}].  Close inspection
reveals that the pursuer orchestrates its actions in such a way to
enforce over time specific relations (linked to the hydrodynamical
cues as discussed in Appendix~\ref{app:mirtail}) between the bearing angles,
namely $\Phi_e\!=\!-\Phi_p$ for mirroring and $\Phi_e\!=\!-\Phi_p
+\pi$ for tailgating [Fig.~\ref{fig:3}(a)]. Due to the aforementioned
$180^o$ ambiguities, the pursuer cannot discern mirroring and
tailgating just on the basis of instantaneous hydrodynamical cues:
The strategy  chosen depends on initial conditions and
hydrodynamic interactions [as, e.g., in Fig.~\ref{fig:3}(b)].
The two strategies emerge from the same policy in
  response to partial observability and can be analytically described,
\textcolor{black}{as detailed in Appendix~\ref{app:mirtail} and briefly summarized in the following}. On neglecting hydrodynamical interactions in
  Eqs.~(\ref{eq:x}) and (\ref{eq:th}), we can derive the equations for the
  separation and bearing angle \cite{belkhouche2007parallel}.  By
  imposing that the pursuer follows either mirroring or tailgating,
  such equations read 
\begin{eqnarray}
\dot{R}&=& -(v_p \pm v_e) \cos\Phi_e \label{eq:R}\\
\dot{\Phi}_e &=&\Omega_e -R^{-1}\; (v_p\mp v_e) \sin\Phi_e \label{eq:phi}\,,
\end{eqnarray}
with $\pm$ for mirroring/tailgating. Equation~(\ref{eq:R}) shows that
tailgating is doomed to fail when $v_p=v_e$ as $\dot{R}=0$, while for
$v_p>v_e$ it becomes an efficient strategy as the dynamics
\eqref{eq:phi} leads to $\Phi_e\!\to\! 0$ for small enough distances,
and \eqref{eq:R} implies $\dot{R}<0$.  Mirroring remains effective
also for $v_p\!=\!v_e$ (and $\Omega_e$ random) as it essentially maps
the pursue into a first hitting problem for a random search with
dimensionality reduction \cite{adam1968reduction}. Tests with RL and
the full dynamics [Eqs.~(\ref{eq:x}) and (\ref{eq:th})] for
$v_p\!=\!v_e$ confirmed the scenario.

\subsubsection{Comparison with a heuristic strategy based on visual cues adapted to partial information \label{sec:baseline}}
\begin{figure}[h!]
\includegraphics[width=0.6\textwidth]{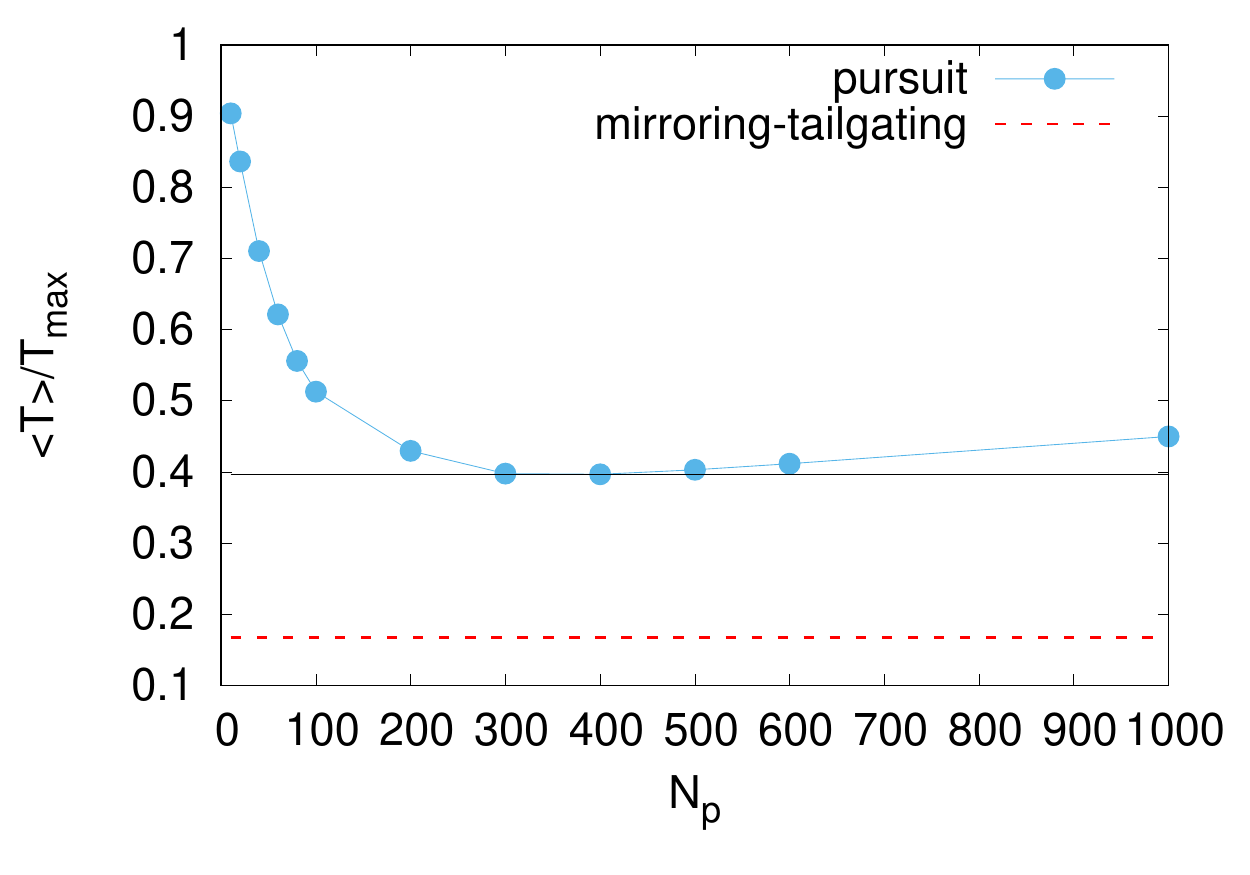}
\caption{\textcolor{black}{(Color online) $\langle T\rangle/T_{max}$ as
    a function of the persistency $N_p$ (circles) of the randomized
    pure pursuit strategy described in text. The black line shows the
    normalized episode duration for the optimal $N_p$, while the red
    dashed line shows the average value obtained with the mirroring-tailgating
    strategy obtained from Fig.~\ref{fig:2}(a).} \label{fig:baseline} }
\end{figure}

It is interesting to compare the pursuit policies discovered by RL
against well-established visual pursuit strategies based on the
knowledge of the line of sight with the target
\cite{nahin2012chases,belkhouche2007parallel}.  Mirroring bears some
similarities with {\it parallel navigation}, where the line-of-sight
direction is kept constant with respect to an inertial frame of
reference, a strategy that appears to be applied by dragonflies
\cite{olberg2000prey}.  Tailgating resembles {\it pure pursuit}, where
heading is constantly directed toward the line of sight (zero bearing
angle), as bats or some fishes appear to do
\cite{chiu2010effects,lanchester1975pursuit}. \textcolor{black}{Such
  strategies cannot be directly implemented here because of the
  $180^{o}$ ambiguities inherent to perceiving only the
  gradients. However, we can introduce a heuristic strategy in the
  form of a randomized pure pursuit: The pursuer heads either toward
  the evader ($\Phi_e\!=\!0$) or to its ``image'' ($\Phi_e\!=\!\pi$)
  with equal probability with some persistency in time.  As a limiting
  case, it could randomly choose its target once for all at the
  beginning, in which case it is bound to fail half of the times so
  that $\langle T\rangle/T_{max}>1/2$; however, the pursuer may
  instead randomly choose either targets every $N_p$ decision times
  (we call $N_p$ persistency). By scanning $\langle T\rangle/T_{max}$
  as a function of $N_p$, at $N_p\approx 400$ we numerically found the
  minimum $\langle T\rangle/T_{max}\approx 0.4$ [see
  Fig.~\ref{fig:baseline} and dashed line in Fig.~\ref{fig:2}(a)] which
  is slightly more than twice the value obtained with the
  mirroring-tailgating strategy.  The policy discovered by RL clearly
  outperforms the randomized pure pursuit offering a more efficient
  way to overcome the ambiguities due to the partial information
  provided by the hydrodynamic cues.}
  
\subsection{Evader strategies: Hydrodynamic defense and linear flights}
In its training phase, the evader learns to contrast mirroring and
tailgating. As for the latter, it finds a way to exploit hydrodynamics
[Fig.~\ref{fig:2}(h)].  In many episodes of this kind, the pursuer
approaches its opponent from behind with small bearing angle
(tailgating).  The evader reacts by placing itself in a position
relative to its predator such that its backward push cancels the speed
advantage of the pursuer and keeps it at bay at a fixed distance
(supplementary movie1 displays the pursuer's trajectory in the frame of
reference of the evader). In principle, near-field
  corrections to the force dipole could modify the hydrodynamic
  defense, and it would be interesting to explore this aspect when
  focusing on specific microswimmers.  Another strategy adopted by
the evader takes the form of an almost linear escape trajectory
[Figs.~\ref{fig:2}(i) and 2(j)]. As shown in Fig.~\ref{fig:2}(d), this is not
always successful as, via mirroring, the pursuer can intercept the
evader to a {\it rendezvous} point by performing a long smooth
arc. Such arcs correspond to adjusting the axis of mirroring in the
course of time.  However, either by making such {\it rendezvous}
point very far [Fig.~\ref{fig:2}(j)] or by exploiting hydrodynamics and
turns on close encounters [Fig.~\ref{fig:2}(i)], the evader can
consistently make its evasion strategies quite efficient.

\subsection{Refining strategies}
As training proceeds both agents learn more complex strategies in
response to the ones described above. We now briefly discuss some
examples that stand out because of their repeated occurrence and
explainability.  Interestingly, the pursuer discovers different ways
to contrast the hydrodynamic defense of its opponent
[Figs.~\ref{fig:2}(e) and (f), see also supplementary
  movie2]. Remarkably, the evader learns to devise diverse winning
manouvers as in Fig.~\ref{fig:2}(k), which consist in linear escapes
and turnings which make the predator switching from mirroring to
tailgating before capture (see supplementary movie3).  The evader also
discovers that twirling can trap the pursuer [Fig.~\ref{fig:2}(l) and 2(m)] in
a looping motion induced by its own mirroring-tailgating
strategy. Trapping is not always successful though
[Fig.~\ref{fig:2}(g)].  With small variations, the basic strategic
patterns discussed above are found also with different parameter
choices and will be reported elsewhere.

\section{Conclusions\label{sec:conclusions}}
In this study, we have shown how microswimmers can discover complex
strategies to pursue and evade from each other, even if endowed with
limited maneuvering ability and inherently equivocal information about
their relative position and orientation. Our study presents a novel
game-theoretic approach to pursuit and evasion in an aquatic
microenvironment.  We expect it to spur further research on the use of
reinforcement learning algorithms to rationalize observed
prey-predator interactions in more general contexts
\cite{hein2020algorithmic}. \textcolor{black}{Owing to the simplicity of
  our model we have been able to analytically describe  some of the
  strategies discovered by RL and show why they are effective in
  overcoming partial observability: For instance, mirroring and tailgating allow to reduce the
  dimensionality of the search by mapping the search into a first hitting problem. In this respect it would
  be interesting to study a three dimensional version of the problem
  to understand which dimensionality reduction could emerge in that
  case and if it can be still reduced to a one-dimensional hitting
  time problem.}

The present model can be easily generalized to ellipsoidal swimmers,
by adding to Eq.~(\ref{eq:th}) rotation by the strain-rate tensor. It
can also be extended to ``pullers'' as well as to other specific
microswimmers by including the appropriate near-field
hydrodynamics. \textcolor{black}{Indeed while, e.g., mirroring and
  tailgating are expected to maintain their efficiency in the far
  field, the pursuer policy may need some refinement in the near-field
  in order to account for more complex hydrodynamic interactions and
  near-field corrections would also modify the response of the evader
  (e.g., the kind of hydrodynamic defense it can develop).}  With
suitable modifications of the hydrodynamics, the approach that we
developed here can be used to train underwater robots which can sense
the hydrodynamic fields with bioinspired
mechanosensors\cite{kottapalli2015soft,free2020bioinspired} and thus
accomplish complex tasks -- for instance, artificial fishes that
imitate escape responses \cite{marchese2014autonomous}.  Here we did
not discuss the effect of external flows and boundaries. Preliminary
results in a circular arena  confirm that the agents
can learn to exploit hydrodynamics to perform their pursue/evasion
tasks in spite of the confounding cues and complex dynamics arising
from the presence of the walls.

Exciting and formidable challenges still lie ahead, and among them
stands out the emergence of collective pursue strategies like
wolf-packing, and collective escape responses such as hydrodynamic
cloaking~\cite{mirzakhanloo2020active}.
  
\begin{acknowledgments}
We thanks Simone Pigolotti for useful comments on the manuscript, and
Xu Zhuoqun for a very careful reading of our manuscript.
F.B. acknowledges hospitality from ICTP. AC has received funding from
the European Union’s Horizon 2020 research and innovation program
under  Marie Skłodowska-Curie Grant No. N956457. This work
received funding from the European Research Council (ERC) under the
European Union’s Horizon 2020 research and innovation programme (Grant
No. 882340).
\end{acknowledgments}

\appendix
\section{Force-dipole hydrodynamic fields\label{app:stokes}}
\textcolor{black}{As described in Sec.~\ref{sec:modelSwimmers},} we
model the agents as two swimming discoids which generate a force
dipole, in the sequel we detail the hydrodynamic fields, which enter
the dynamics of the agents [see Eqs.~(\ref{eq:x}) and (\ref{eq:th})],
generated by a force dipole in two dimensions.

We start considering a Stokeslet, i.e., the fundamental solution of
the Stokes equation for a point force, $\bm F=F\bm n$, which, for the
sake of simplicity, we locate in the origin, and thus solving the
equation
\begin{equation}
\nu \Delta \bm u-\bm \nabla p=\bm F\delta(\bm x)\,,
\label{eq:stokes}
\end{equation}
where $\bm u$ and $p$ are the velocity and pressure field, and $\nu$
the fluid viscosity.  The fundamental solution to
Eq.~(\ref{eq:stokes}) in two dimensions is
\begin{equation}
  u_i(\bm x)=G_{ij}(\bm x) F_j\,,
  \label{eq:sol1}
\end{equation}
where $G$ is the Green function: 
\begin{equation}
  G_{ij}(\bm x)= \frac{1}{4\pi\nu} \left[ -\delta_{ij} \ln \left(\frac{|\bm x|}{L}\right)+\frac{x_ix_j}{|\bm x|^2}\right]
\end{equation}
with $L$ being an arbitrary length. The pressure field takes the form
$p(\bm x)=\bm F\cdot \bm x/(4\pi |\bm x|^3)+p_\infty$, with $p_\infty$ a constant.

Considering two point forces $\bm F^{\pm}=\pm F \bm n$ located in
$\bm x^{\pm}=\pm \epsilon \bm n$, with $\epsilon\ll 1$ and using
Eq.~(\ref{eq:sol1}), we can express the velocity field generated by this couple  as
\begin{equation}
  \bm u(\bm x)= G_{ij}(\bm x-\bm x^+) F^+_j+G_{ij}(\bm x-\bm x^+) F^-_j \simeq -2F n_k \partial_k G_{ij}(\bm x) n_j
\end{equation}
where $F^+_i=-F^-_i=Fn_i$ and we retained only the
first order, to obtain an expression which well approximates the
velocity field for large distances $|\bm x|\gg \epsilon$. Working out
the algebra yields:
\begin{equation}
  \bm u(\bm x)= \frac{D}{|\bm x|} \left[2\left(\frac{\bm n\cdot\bm x}{|\bm x|}\right)^2-1\right] \frac{\bm x}{|\bm x|} = \frac{D}{|\bm x|} \cos(2\phi-2\theta) (\cos\phi,\sin\phi)
  \label{eq:vel}
\end{equation}
where  $D=F\epsilon/(2\pi\nu)$  measures the dipole intensity ($D>0$ corresponding to pushers and $D<0$ to pullers \cite{lauga2009hydrodynamics}) and, in the second equality, $\bm x=|\bm x|(\cos\phi,\sin\phi)$ and $\bm n=\bm n(\theta)=(\cos\theta,\sin\theta)$.
Notice that the velocity (\ref{eq:vel}) can equivalently be derived as
$\bm u=(\partial_y \Psi,-\partial_x \Psi)$ where $\Psi$ is the stream function
which can be written as
\begin{equation}
\Psi(\bm x)=\frac{D}{2}\sin(2\phi-2\theta)\,.
\end{equation}

\begin{figure}[h!]
\includegraphics[width=0.6\linewidth]{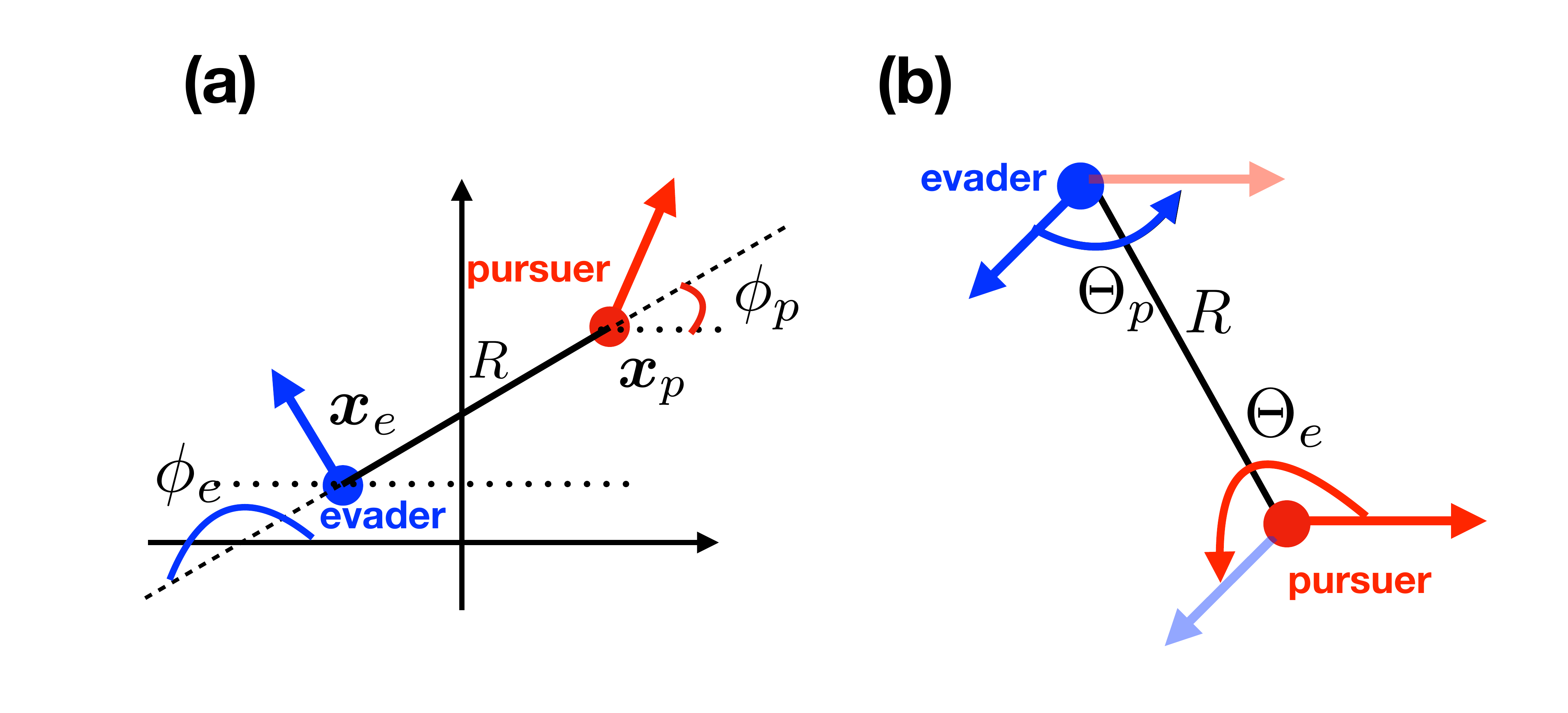}
 \caption{(Color online) \label{fig:geom} Description of relevant angles entering the agent dynamics. (a) In a fixed frame of reference it is shown the angular position of the pursuer $\phi_p$ and of the evader $\phi_e$, heading angles are shown in Fig.~\ref{fig:1}(b). (b) We show the relative heading angles $\Theta_e=\theta_e-\theta_p$ and $\Theta_p=\theta_p-\theta_e$, needed together with the bearing angles [Fig.~\ref{fig:1}(d)] to express the velocity gradients in the pursuer and evader frame of reference, respectively. }
\end{figure}

The velocity field due to agent $\beta$ and advecting agent $\alpha$
[see Eq.~(\ref{eq:x}] is simply obtained from Eq.~(\ref{eq:vel})
substituting $\phi=\phi_\alpha$ and $\theta=\theta_\beta$, where
$\theta_\beta$ are the heading directions shown in Fig.~\ref{fig:1}(c), and $\phi_\alpha$ the angular position with respect to a fixed
frame of reference shown in Fig.~\ref{fig:geom}(a).
The vorticity field due to agent $\beta$ and rotating the heading orientation $\theta_\alpha$ of agent $\alpha$ instead
can be easily derived to be:
\begin{equation}
\omega^{(\beta)}=\frac{2D_\beta}{R^2}\sin(2\phi_\alpha-2\theta_\beta)=
\frac{2D_\beta}{R^2}\sin(2\phi_\beta-2\theta_\beta)
\end{equation}
 with $R=|\bm
x_\alpha-\bm x_\beta|$
and the second equality stemming from
$\phi_\beta=\phi_\alpha+\pi$ [Fig.~\ref{fig:geom}(a)].

\section{Choice of observables and features \label{app:features}}
Equations~(\ref{eq:vort})-(\ref{eq:shear}) express the gradients in
the frame of reference the observing agent. The three independent
components of the gradients $\omega,\mathcal{L}$ and $\mathcal{S}$ can be
mapped one to one onto the space of the following quantities
$o=\{\omega,\hat{R},\gamma\}$, which we assume are observables. Here,
$\omega$ is the vorticity itself, which combines information about the
agent distance and about $\sin(2\Phi-2\Theta)$.  The other two
quantities can easily obtained combining the expression of the
longitudinal and shear strain, as
$\hat{R}\!=\!(\mathcal{L}^2\!+\!\mathcal{S}^2)^{1/2}\propto 1/R^2$ and
$\gamma=\mathrm{arctan2}(\mathcal{S}/\mathcal{L})$.  The observations
$o$ offer partial information about the other agent due to the $180^o$
ambiguities in $\Theta$ and $\Phi$ discussed in
Sec.~\ref{sec:modelCues}. The policy is parameterized by the features
$\mathcal{F}(o)$, i.e., functions of the observables. Notice that even
if $o$ was precisely identifying the reciprocal position of the
agents, in order to have access to the full space of possible
policies, the features should be chosen as a complete functional basis
of the observation, which is not practicable if not using deep
reinforcement learning techniques, an option that we did not adopt to
have a better understanding of the discovered
policies. \textcolor{black}{So we encode} the observations $o$ by using
a set of only $N_F$ features and, consequently, the agents must decide
their actions in condition of \textit{partial observability}
\cite{sutton2018reinforcement,jaakkola1995reinforcement}).

We tested different choices of the features and the results presented
in Fig.~\ref{fig:2} correspond to the choice of the $N_F=13$ features 
summarized in Table~\ref{tab:features}.
\begin{table}[h!]
\centering
\begin{tabular}{|l|}
\hline
$\mathcal{F}_1(o_t)= \hat{R}(t)$\\
\hline
$\mathcal{F}_2(o_t)= \sin(\gamma(t))$\\
\hline
$\mathcal{F}_3(o_t)= \cos(\gamma(t))$\\
\hline
$\mathcal{F}_4(o_t)= \sin(2\,\gamma(t))$\\
\hline
$\mathcal{F}_5(o_t)= \cos(2\,\gamma(t))$\\
\hline
$\mathcal{F}_6(o_t)= \omega(t)$\\
\hline
$\mathcal{F}_i(o_t)=(1-\mu)\;\mathcal{F}_i(t-\tau)+\mu\;\mathcal{F}_{i-6}(t-\tau) \quad \mathrm{for} \quad i=7,12$\\
\hline
$\mathcal{F}_{13}(o_t)=1$\\
\hline
\end{tabular}\caption{\label{tab:features} Implemented Features. Note that
  features $7-12$ provide some memory of the values of the previous
  features, in the implementation we chose $\mu=0.3$  to
  retain some memory about the last $2-4$ past observations
  approximately).  }
\end{table}
While the first six features
are clearly related to the information that can be extracted from
gradients, the features $i=7,12$ are introduced to provide the agents
with some memory, which may mitigate some of the aforementioned
ambiguities. Finally, the $13^{th}$ feature is unrelated to the
gradients and it is chosen to allow the agents to adopt strategies
independent of the percepts.

We remark that the mirroring and tailgating strategies discussed in
Sec.~\ref{sec:mt} can be obtained also removing memory (i.e., features from $7$
to $12$) while they seem to crucially depend on features 4 and 5
which, as explained before, are derived from the strain components
Eqs.~\eqref{eq:long} and \eqref{eq:shear}. These two features are
clearly related to mirroring and tailgating strategies (see also
Appendix~\ref{app:mirtail}). Indeed, by removing them, such basic
strategies are lost.  In order to assess the robustness of our
results, we tested the algorithm with different choices of
features. Specifically, we tried using powers (both positive and
negative) of $\hat R$, higher harmonics of $\gamma$ and various
products of the percepts. For instance, we tried to add the following
features $\cos(3\gamma)$, $\sin(4\gamma)$, $\cos(3\gamma)$,
$\cos(4\gamma)$, $\omega \cos(\gamma)/\hat R$, $\omega
\sin(\gamma)/\hat R$, $\omega \sin(2\gamma)/\hat R$, $\omega \cos(2
\gamma)/\hat R$, $\hat R \cos(\gamma)$, $\hat R \sin(\gamma)$, $\hat R
\sin(2\gamma)$, $\hat R \cos(2\gamma)$, $\hat R \sin(3\gamma)$, $\hat
R \cos(3 \gamma)$, $|\omega|$ and $\omega/\hat R$. In a separate batch
of tests, we tried to use $\hat R^2$ and $ \log \hat R$. While it
cannot be excluded that we did miss a specific combination of features
or that we did not run enough tests, the strategies emerging from
these additional trials were qualitative equivalent to those presented
in Fig.~\ref{fig:2}.

\section{Analytical description of mirroring and tailgating strategies\label{app:mirtail}}
In this Appendix we discuss the mirroring [Fig.~\ref{fig:2}(b)] and tailgating ([Fig.~\ref{fig:2}(c)] strategies and derive Eqs.~(\ref{eq:R}) and (\ref{eq:phi}).

First we recall the definitions [see also Figs.~\ref{fig:1}(c) and 1(d)  and
Figs.~\ref{fig:geom}(a) and 5(b)] of the relative heading angle
$\Theta_e=\theta_e-\theta_p$, bearing angle from the point of view of
the pursuer, $\Phi_{e}=\phi_{e}-\theta_{p}$, and of the evader,
$\Phi_{p}=\phi_{p}-\theta_{e}$; moreover,  we
recall that $\phi_{e}=\phi_{p}+\pi$ [see Fig.~\ref{fig:geom}(a)].  As discussed in Sec.~\ref{sec:mt} in
these strategies the pursuer chooses its actions in such a way to
approximatively keep the following relations between the two bearing
angles:
\begin{equation}\label{condhsm}
\begin{cases}
\Phi_p= -\Phi_e \phantom{+ \pi}\;\;\qquad \mathrm{Mirroring} \\
\Phi_p= -\Phi_e +\pi  \qquad \mathrm{Tailgating}\,. 
\end{cases}
\end{equation}
As discussed in Appendix~\ref{app:features}, one of the key observables available to
the pursuer is the angle $2\Phi_e-\Theta_e$, with simple algebra one can recognize that
\begin{equation}
2\Phi_e-\Theta_e=\Phi_p+\Phi_e-\pi\,,
\end{equation}  
so that we can re-express \eqref{condhsm} as 
\begin{equation} \label{mirrpm}
2\Phi_e-\Theta_e=\Gamma_{\pm} \mod 2\pi
\end{equation}
or equivalently, in the laboratory frame of reference
\begin{equation}
  \theta_p+\theta_e=2\phi_e-\Gamma_{\pm}\,,
  \label{mirrpm2}
\end{equation}
with $\Gamma_+=\pi$ for mirroring and $\Gamma_-=0$ for tailgating
respectively. Note that $(\theta_p+\theta_e)/2$ identifies (for
$v_p=v_e$ exactly, and otherwise approximatively) the axis of symmetry
with respect to which the pursuer trajectory mirrors the one of the
evader.

Notice that, at least in the absence of memory, $\Gamma_\pm$ cannot be
discriminated from observing gradients alone due to the fore-aft
symmetry of the swimming dipole. Therefore, depending on the initial
condition the agent will pick one of two strategies.  Unless the
pursuer can resolve the aforementioned ambiguity, it must learn both
strategies or neither.
In the actual hydrodynamic simulations we have added memory effects
(see features $\mathcal{F}_i$ for $i=7,12$ in
Table~\ref{tab:features}) to possibly allow the agents to break the
fore-aft symmetry in observations. Though tests in the absence of
memory suggest that the way memory was implemented is likely not
sufficient to eliminate such ambiguities. Anyway, we will ignore
possible memory effects in the following.

In order to explore the basic features of such strategies we will
neglect also the hydrodynamic effects (i.e., we will not consider the
effects on the pursuer due to the velocity field induced by the evader)
meaning that we  approximate Eqs.~(\ref{eq:x}) and (\ref{eq:th})  as
\begin{eqnarray}
  \dot{\bm x}_\alpha &=& v_\alpha \bm n(\theta_\alpha)\label{eq:x1} \\
  \dot{\theta}_\alpha &=& \Omega_\alpha \label{eq:th1}\,.
\end{eqnarray}
Moreover, while for the evading agent we assume the dynamics as given
by Eqs.~(\ref{eq:x1}) and (\ref{eq:th1}) with $\Omega_e$ chosen random among
the values $0,\pm \varpi_e$, as in Ref.~\cite{belkhouche2007parallel}
for the pursuer we enforce the constraint (\ref{mirrpm2}) exactly. It
is worth underlying that the above kinematic equations remain valid
also for $\Omega_e$ non random.  Then we can derive -- in polar
coordinates $R=|\bm x_e-\bm x_p|$ and $\Phi_e$ -- the motion of the
pursuer in the pursuer frame of reference.

\textit{Computing $\dot R$:\quad} In order to compute $\dot R$, we
have to project both velocities on the direction ($\phi_{e}$)
connecting the two agents. We can use
Eqs.~(\ref{mirrpm})-(\ref{condhsm}) (i.e., we can either project
velocities onto the pursuer to evader direction of motion or compute
$\dot R$ with the chain rule). This procedure leads to $\dot
R=v_e\cos(\theta_e-\phi_{e})-v_p\cos(\theta_p-\phi_{e})=v_e\cos(\Phi_e-\Gamma_\pm)-v_p\cos(\Phi_e)$,
and thus to $\dot R=-(v_p\pm v_e)\cos(\Phi_e)$.

\textit{Computing $\dot \Phi_e$:\quad} Let us now focus on $\dot
\Phi_e$. By using Eq.~(\ref{mirrpm2}) and that
$\dot{\theta}_e=\Omega_e$ (being $\Omega_e$ the angular velocity
selected by the pursuer), we can  deduce that
$\dot\theta_p=2\dot\phi_{e}-\Omega_e$ which implies that
\begin{equation}\label{bfhwe}
\dot \Phi_e=-\dot \phi_{e}+\Omega_e
\end{equation}
On the other hand, a direct computation yields 
\begin{equation}
  \dot \phi_{e}=\frac{1}{R}[v_e\sin(\theta_e-\phi_{e})-v_p\sin(\theta_p-\phi_{p})]\,.
  \label{eq:phihsdot}
\end{equation}
Now using Eqs.~(\ref{bfhwe}) and (\ref{eq:phihsdot})
and noticing that $\theta_e-\phi_{e}=\phi_{e}-\theta_p-\Gamma_\pm=\Phi_e-\Gamma_\pm$ from (\ref{mirrpm2}), we can deduce that
\begin{equation}
\dot \Phi_e=\Omega_e-\frac{1}{R}[v_e\sin(\Phi_e-\Gamma_\pm)+v_p\sin(\Phi_e)]=\Omega_e-\frac{1}{R}(v_p\mp v_e)\sin(\Phi_e)\,.
\end{equation}

We can then summarize the previous results in the set of equations
\begin{equation}
  \label{eq:relativedyn}
\begin{dcases}
\dot R&=-(v_p\pm v_e)\cos(\Phi_e)\\
\dot\Phi_e&=\Omega_e-\frac{1}{R}(v_p\mp v_e)\sin(\Phi_e)\,,
\end{dcases}
\end{equation}
where we recall the upper sign choice applies to mirroring  and
the lower choice to tailgating. Notice that
Eq.~(\ref{eq:relativedyn}) essentially coincides with Eq.~(17) of
Ref.~\cite{belkhouche2007parallel} but is specialized to the
mirroring/tailgating constraint (\ref{mirrpm2}) on the angles.


\begin{thebibliography}{49}%
\makeatletter
\providecommand \@ifxundefined [1]{%
 \@ifx{#1\undefined}
}%
\providecommand \@ifnum [1]{%
 \ifnum #1\expandafter \@firstoftwo
 \else \expandafter \@secondoftwo
 \fi
}%
\providecommand \@ifx [1]{%
 \ifx #1\expandafter \@firstoftwo
 \else \expandafter \@secondoftwo
 \fi
}%
\providecommand \natexlab [1]{#1}%
\providecommand \enquote  [1]{``#1''}%
\providecommand \bibnamefont  [1]{#1}%
\providecommand \bibfnamefont [1]{#1}%
\providecommand \citenamefont [1]{#1}%
\providecommand \href@noop [0]{\@secondoftwo}%
\providecommand \href [0]{\begingroup \@sanitize@url \@href}%
\providecommand \@href[1]{\@@startlink{#1}\@@href}%
\providecommand \@@href[1]{\endgroup#1\@@endlink}%
\providecommand \@sanitize@url [0]{\catcode `\\12\catcode `\$12\catcode
  `\&12\catcode `\#12\catcode `\^12\catcode `\_12\catcode `\%12\relax}%
\providecommand \@@startlink[1]{}%
\providecommand \@@endlink[0]{}%
\providecommand \url  [0]{\begingroup\@sanitize@url \@url }%
\providecommand \@url [1]{\endgroup\@href {#1}{\urlprefix }}%
\providecommand \urlprefix  [0]{URL }%
\providecommand \Eprint [0]{\href }%
\providecommand \doibase [0]{http://dx.doi.org/}%
\providecommand \selectlanguage [0]{\@gobble}%
\providecommand \bibinfo  [0]{\@secondoftwo}%
\providecommand \bibfield  [0]{\@secondoftwo}%
\providecommand \translation [1]{[#1]}%
\providecommand \BibitemOpen [0]{}%
\providecommand \bibitemStop [0]{}%
\providecommand \bibitemNoStop [0]{.\EOS\space}%
\providecommand \EOS [0]{\spacefactor3000\relax}%
\providecommand \BibitemShut  [1]{\csname bibitem#1\endcsname}%
\let\auto@bib@innerbib\@empty
\bibitem [{\citenamefont {Triantafyllou}\ \emph {et~al.}(2016)\citenamefont
  {Triantafyllou}, \citenamefont {Weymouth},\ and\ \citenamefont
  {Miao}}]{triantafyllou2016biomimetic}%
  \BibitemOpen
  \bibfield  {author} {\bibinfo {author} {\bibfnamefont {M.~S.}\ \bibnamefont
  {Triantafyllou}}, \bibinfo {author} {\bibfnamefont {G.~D.}\ \bibnamefont
  {Weymouth}}, \ and\ \bibinfo {author} {\bibfnamefont {J.}~\bibnamefont
  {Miao}},\ }\bibfield  {title} {\enquote {\bibinfo {title} {Biomimetic
  survival hydrodynamics and flow sensing},}\ }\href@noop {} {\bibfield
  {journal} {\bibinfo  {journal} {Ann. Rev. Fluid Mech.}\ }\textbf {\bibinfo
  {volume} {48}},\ \bibinfo {pages} {1} (\bibinfo {year} {2016})}\BibitemShut
  {NoStop}%
\bibitem [{\citenamefont {Takagi}\ and\ \citenamefont
  {Hartline}(2018)}]{takagi2018directional}%
  \BibitemOpen
  \bibfield  {author} {\bibinfo {author} {\bibfnamefont {D.}~\bibnamefont
  {Takagi}}\ and\ \bibinfo {author} {\bibfnamefont {D.~K.}\ \bibnamefont
  {Hartline}},\ }\bibfield  {title} {\enquote {\bibinfo {title} {Directional
  hydrodynamic sensing by free-swimming organisms},}\ }\href@noop {} {\bibfield
   {journal} {\bibinfo  {journal} {Bull. Math. Biol.}\ }\textbf {\bibinfo
  {volume} {80}},\ \bibinfo {pages} {215} (\bibinfo {year} {2018})}\BibitemShut
  {NoStop}%
\bibitem [{\citenamefont {Tuttle}\ \emph {et~al.}(2019)\citenamefont {Tuttle},
  \citenamefont {Robinson}, \citenamefont {Takagi}, \citenamefont {Strickler},
  \citenamefont {Lenz},\ and\ \citenamefont {Hartline}}]{tuttle2019going}%
  \BibitemOpen
  \bibfield  {author} {\bibinfo {author} {\bibfnamefont {L.~J.}\ \bibnamefont
  {Tuttle}}, \bibinfo {author} {\bibfnamefont {H.~E.}\ \bibnamefont
  {Robinson}}, \bibinfo {author} {\bibfnamefont {D.}~\bibnamefont {Takagi}},
  \bibinfo {author} {\bibfnamefont {J~R.}\ \bibnamefont {Strickler}}, \bibinfo
  {author} {\bibfnamefont {P.~H.}\ \bibnamefont {Lenz}}, \ and\ \bibinfo
  {author} {\bibfnamefont {D.~K.}\ \bibnamefont {Hartline}},\ }\bibfield
  {title} {\enquote {\bibinfo {title} {Going with the flow: hydrodynamic cues
  trigger directed escapes from a stalking predator},}\ }\href@noop {}
  {\bibfield  {journal} {\bibinfo  {journal} {J. Royal Soc. Interface}\
  }\textbf {\bibinfo {volume} {16}},\ \bibinfo {pages} {20180776} (\bibinfo
  {year} {2019})}\BibitemShut {NoStop}%
\bibitem [{\citenamefont {Lloyd}\ \emph {et~al.}(2018)\citenamefont {Lloyd},
  \citenamefont {Olive}, \citenamefont {Stahl}, \citenamefont {Jaggard},
  \citenamefont {Amaral}, \citenamefont {Dubou{\'e}},\ and\ \citenamefont
  {Keene}}]{lloyd2018evolutionary}%
  \BibitemOpen
  \bibfield  {author} {\bibinfo {author} {\bibfnamefont {E.}~\bibnamefont
  {Lloyd}}, \bibinfo {author} {\bibfnamefont {C.}~\bibnamefont {Olive}},
  \bibinfo {author} {\bibfnamefont {B.~A.}\ \bibnamefont {Stahl}}, \bibinfo
  {author} {\bibfnamefont {J.~B.}\ \bibnamefont {Jaggard}}, \bibinfo {author}
  {\bibfnamefont {P.}~\bibnamefont {Amaral}}, \bibinfo {author} {\bibfnamefont
  {E.~R.}\ \bibnamefont {Dubou{\'e}}}, \ and\ \bibinfo {author} {\bibfnamefont
  {A.~C.}\ \bibnamefont {Keene}},\ }\bibfield  {title} {\enquote {\bibinfo
  {title} {Evolutionary shift towards lateral line dependent prey capture
  behavior in the blind mexican cavefish},}\ }\href@noop {} {\bibfield
  {journal} {\bibinfo  {journal} {Develop. Biol.}\ }\textbf {\bibinfo {volume}
  {441}},\ \bibinfo {pages} {328} (\bibinfo {year} {2018})}\BibitemShut
  {NoStop}%
\bibitem [{\citenamefont {Montgomery}\ \emph {et~al.}(1997)\citenamefont
  {Montgomery}, \citenamefont {Baker},\ and\ \citenamefont
  {Carton}}]{montgomery1997lateral}%
  \BibitemOpen
  \bibfield  {author} {\bibinfo {author} {\bibfnamefont {J.~C.}\ \bibnamefont
  {Montgomery}}, \bibinfo {author} {\bibfnamefont {C.~F.}\ \bibnamefont
  {Baker}}, \ and\ \bibinfo {author} {\bibfnamefont {A.~G.}\ \bibnamefont
  {Carton}},\ }\bibfield  {title} {\enquote {\bibinfo {title} {The lateral line
  can mediate rheotaxis in fish},}\ }\href@noop {} {\bibfield  {journal}
  {\bibinfo  {journal} {Nature}\ }\textbf {\bibinfo {volume} {389}},\ \bibinfo
  {pages} {960} (\bibinfo {year} {1997})}\BibitemShut {NoStop}%
\bibitem [{\citenamefont {Bleckmann}\ and\ \citenamefont
  {Zelick}(2009)}]{bleckmann2009lateral}%
  \BibitemOpen
  \bibfield  {author} {\bibinfo {author} {\bibfnamefont {H.}~\bibnamefont
  {Bleckmann}}\ and\ \bibinfo {author} {\bibfnamefont {R.}~\bibnamefont
  {Zelick}},\ }\bibfield  {title} {\enquote {\bibinfo {title} {Lateral line
  system of fish},}\ }\href@noop {} {\bibfield  {journal} {\bibinfo  {journal}
  {Integr. Zool.}\ }\textbf {\bibinfo {volume} {4}},\ \bibinfo {pages} {13}
  (\bibinfo {year} {2009})}\BibitemShut {NoStop}%
\bibitem [{\citenamefont {Kanter}\ and\ \citenamefont
  {Coombs}(2003)}]{kanter2003rheotaxis}%
  \BibitemOpen
  \bibfield  {author} {\bibinfo {author} {\bibfnamefont {M.~J.}\ \bibnamefont
  {Kanter}}\ and\ \bibinfo {author} {\bibfnamefont {S.}~\bibnamefont
  {Coombs}},\ }\bibfield  {title} {\enquote {\bibinfo {title} {Rheotaxis and
  prey detection in uniform currents by lake michigan mottled sculpin (cottus
  bairdi)},}\ }\href@noop {} {\bibfield  {journal} {\bibinfo  {journal} {J.
  Experm. Biol.}\ }\textbf {\bibinfo {volume} {206}},\ \bibinfo {pages} {59}
  (\bibinfo {year} {2003})}\BibitemShut {NoStop}%
\bibitem [{\citenamefont {Ki{\o}rboe}\ and\ \citenamefont
  {Visser}(1999)}]{kiorboe1999predator}%
  \BibitemOpen
  \bibfield  {author} {\bibinfo {author} {\bibfnamefont {T.}~\bibnamefont
  {Ki{\o}rboe}}\ and\ \bibinfo {author} {\bibfnamefont {A.~W.}\ \bibnamefont
  {Visser}},\ }\bibfield  {title} {\enquote {\bibinfo {title} {Predator and
  prey perception in copepods due to hydromechanical signals},}\ }\href@noop {}
  {\bibfield  {journal} {\bibinfo  {journal} {Mar. Ecol. Progr. Ser.}\ }\textbf
  {\bibinfo {volume} {179}},\ \bibinfo {pages} {81} (\bibinfo {year}
  {1999})}\BibitemShut {NoStop}%
\bibitem [{\citenamefont {Doall}\ \emph {et~al.}(2002)\citenamefont {Doall},
  \citenamefont {Strickler}, \citenamefont {Fields},\ and\ \citenamefont
  {Yen}}]{doall2002mapping}%
  \BibitemOpen
  \bibfield  {author} {\bibinfo {author} {\bibfnamefont {M.}~\bibnamefont
  {Doall}}, \bibinfo {author} {\bibfnamefont {J.}~\bibnamefont {Strickler}},
  \bibinfo {author} {\bibfnamefont {D.}~\bibnamefont {Fields}}, \ and\ \bibinfo
  {author} {\bibfnamefont {J.}~\bibnamefont {Yen}},\ }\bibfield  {title}
  {\enquote {\bibinfo {title} {Mapping the free-swimming attack volume of a
  planktonic copepod, euchaeta rimana},}\ }\href@noop {} {\bibfield  {journal}
  {\bibinfo  {journal} {Mar. Biol.}\ }\textbf {\bibinfo {volume} {140}},\
  \bibinfo {pages} {871} (\bibinfo {year} {2002})}\BibitemShut {NoStop}%
\bibitem [{\citenamefont {Happel}\ and\ \citenamefont
  {Brenner}(2012)}]{happel2012low}%
  \BibitemOpen
  \bibfield  {author} {\bibinfo {author} {\bibfnamefont {J.}~\bibnamefont
  {Happel}}\ and\ \bibinfo {author} {\bibfnamefont {H.}~\bibnamefont
  {Brenner}},\ }\href@noop {} {\emph {\bibinfo {title} {Low Reynolds Number
  Hydrodynamics: With Special Applications to Particulate Media}}},\
  Vol.~\bibinfo {volume} {1}\ (\bibinfo  {publisher} {Springer Science \&
  Business Media, New York},\ \bibinfo {year} {2012})\BibitemShut {NoStop}%
\bibitem [{\citenamefont {Hein}\ \emph {et~al.}(2020)\citenamefont {Hein},
  \citenamefont {Altshuler}, \citenamefont {Cade}, \citenamefont {Liao},
  \citenamefont {Martin},\ and\ \citenamefont {Taylor}}]{hein2020algorithmic}%
  \BibitemOpen
  \bibfield  {author} {\bibinfo {author} {\bibfnamefont {A.~M.}\ \bibnamefont
  {Hein}}, \bibinfo {author} {\bibfnamefont {D.~L.}\ \bibnamefont {Altshuler}},
  \bibinfo {author} {\bibfnamefont {D.~E.}\ \bibnamefont {Cade}}, \bibinfo
  {author} {\bibfnamefont {J.~C.}\ \bibnamefont {Liao}}, \bibinfo {author}
  {\bibfnamefont {B.~T.}\ \bibnamefont {Martin}}, \ and\ \bibinfo {author}
  {\bibfnamefont {G.~K.}\ \bibnamefont {Taylor}},\ }\bibfield  {title}
  {\enquote {\bibinfo {title} {An algorithmic approach to natural behavior},}\
  }\href@noop {} {\bibfield  {journal} {\bibinfo  {journal} {Curr. Biol.}\
  }\textbf {\bibinfo {volume} {30}},\ \bibinfo {pages} {R663} (\bibinfo {year}
  {2020})}\BibitemShut {NoStop}%
\bibitem [{\citenamefont {Domenici}\ \emph {et~al.}(2011)\citenamefont
  {Domenici}, \citenamefont {Blagburn},\ and\ \citenamefont
  {Bacon}}]{domenici2011animal}%
  \BibitemOpen
  \bibfield  {author} {\bibinfo {author} {\bibfnamefont {P.}~\bibnamefont
  {Domenici}}, \bibinfo {author} {\bibfnamefont {J.~M.}\ \bibnamefont
  {Blagburn}}, \ and\ \bibinfo {author} {\bibfnamefont {J.~P.}\ \bibnamefont
  {Bacon}},\ }\bibfield  {title} {\enquote {\bibinfo {title} {Animal escapology
  i: Theoretical issues and emerging trends in escape trajectories},}\
  }\href@noop {} {\bibfield  {journal} {\bibinfo  {journal} {J. Exp. Biol.}\
  }\textbf {\bibinfo {volume} {214}},\ \bibinfo {pages} {2463} (\bibinfo {year}
  {2011})}\BibitemShut {NoStop}%
\bibitem [{\citenamefont {Nahin}(2012)}]{nahin2012chases}%
  \BibitemOpen
  \bibfield  {author} {\bibinfo {author} {\bibfnamefont {P.~J}\ \bibnamefont
  {Nahin}},\ }\href@noop {} {\emph {\bibinfo {title} {Chases and Escapes: The
  Mathematics of Pursuit and Evasion}}}\ (\bibinfo  {publisher} {Princeton
  University Press, Princeton, NJ},\ \bibinfo {year} {2012})\BibitemShut {NoStop}%
\bibitem [{\citenamefont {Hofbauer}\ and\ \citenamefont
  {Sigmund}(1998)}]{hofbauer1998evolutionary}%
  \BibitemOpen
  \bibfield  {author} {\bibinfo {author} {\bibfnamefont {J.}~\bibnamefont
  {Hofbauer}}\ and\ \bibinfo {author} {\bibfnamefont {K.}~\bibnamefont
  {Sigmund}},\ }\href@noop {} {\emph {\bibinfo {title} {Evolutionary Games and
  Population Dynamics}}}\ (\bibinfo  {publisher} {Cambridge University Press, Cambridge, UK},\
  \bibinfo {year} {1998})\BibitemShut {NoStop}%
\bibitem [{\citenamefont {Sutton}\ and\ \citenamefont
  {Barto}(2018)}]{sutton2018reinforcement}%
  \BibitemOpen
  \bibfield  {author} {\bibinfo {author} {\bibfnamefont {R.~S.}\ \bibnamefont
  {Sutton}}\ and\ \bibinfo {author} {\bibfnamefont {A.~G.}\ \bibnamefont
  {Barto}},\ }\href@noop {} {\emph {\bibinfo {title} {Reinforcement learning:
  An introduction}}}\ (\bibinfo  {publisher} {MIT Press, Cambridge, MA},\ \bibinfo {year}
  {2018})\BibitemShut {NoStop}%
\bibitem [{\citenamefont {Baker}\ \emph {et~al.}(2019)\citenamefont {Baker},
  \citenamefont {Kanitscheider}, \citenamefont {Markov}, \citenamefont {Wu},
  \citenamefont {Powell}, \citenamefont {McGrew},\ and\ \citenamefont
  {Mordatch}}]{baker2019emergent}%
  \BibitemOpen
  \bibfield  {author} {\bibinfo {author} {\bibfnamefont {B.}~\bibnamefont
  {Baker}}, \bibinfo {author} {\bibfnamefont {I.}~\bibnamefont
  {Kanitscheider}}, \bibinfo {author} {\bibfnamefont {T.}~\bibnamefont
  {Markov}}, \bibinfo {author} {\bibfnamefont {Y.}~\bibnamefont {Wu}}, \bibinfo
  {author} {\bibfnamefont {G.}~\bibnamefont {Powell}}, \bibinfo {author}
  {\bibfnamefont {B.}~\bibnamefont {McGrew}}, \ and\ \bibinfo {author}
  {\bibfnamefont {I.}~\bibnamefont {Mordatch}},\ }\bibfield  {title} {\enquote
  {\bibinfo {title} {Emergent tool use from multi-agent autocurricula},}\ }in\
  \href@noop {} {\emph {\bibinfo {booktitle} {Proceeding of the International Conference on
  Learning Representations}}}\ (\bibinfo {year} {2019})\BibitemShut {NoStop}%
\bibitem [{\citenamefont {Chen}\ \emph {et~al.}(2020)\citenamefont {Chen},
  \citenamefont {Song}, \citenamefont {Lipson},\ and\ \citenamefont
  {Vondrick}}]{chen2020visual}%
  \BibitemOpen
  \bibfield  {author} {\bibinfo {author} {\bibfnamefont {B.}~\bibnamefont
  {Chen}}, \bibinfo {author} {\bibfnamefont {S.}~\bibnamefont {Song}}, \bibinfo
  {author} {\bibfnamefont {H.}~\bibnamefont {Lipson}}, \ and\ \bibinfo {author}
  {\bibfnamefont {C.}~\bibnamefont {Vondrick}},\ }\bibfield  {title} {\enquote
  {\bibinfo {title} {Visual hide and seek},}\ }in\ \href@noop {} {\emph
  {\bibinfo {booktitle} {Artificial Life Conference Proceedings}}}\ (\bibinfo
  {organization} {MIT Press, Cambridge, MA},\ \bibinfo {year} {2020})\ pp.\ \bibinfo {pages}
  {645--655}\BibitemShut {NoStop}%
\bibitem [{\citenamefont {Biferale}\ \emph {et~al.}(2019)\citenamefont
  {Biferale}, \citenamefont {Bonaccorso}, \citenamefont {Buzzicotti},
  \citenamefont {Clark Di~Leoni},\ and\ \citenamefont
  {Gustavsson}}]{biferale2019zermelo}%
  \BibitemOpen
  \bibfield  {author} {\bibinfo {author} {\bibfnamefont {L.}~\bibnamefont
  {Biferale}}, \bibinfo {author} {\bibfnamefont {F.}~\bibnamefont
  {Bonaccorso}}, \bibinfo {author} {\bibfnamefont {M.}~\bibnamefont
  {Buzzicotti}}, \bibinfo {author} {\bibfnamefont {P.}~\bibnamefont {Clark
  Di~Leoni}}, \ and\ \bibinfo {author} {\bibfnamefont {K.}~\bibnamefont
  {Gustavsson}},\ }\bibfield  {title} {\enquote {\bibinfo {title} {Zermelo’s
  problem: Optimal point-to-point navigation in 2d turbulent flows using
  reinforcement learning},}\ }\href@noop {} {\bibfield  {journal} {\bibinfo
  {journal} {Chaos}\ }\textbf {\bibinfo {volume} {29}},\ \bibinfo {pages}
  {103138} (\bibinfo {year} {2019})}\BibitemShut {NoStop}%
\bibitem [{\citenamefont {Alageshan}\ \emph {et~al.}(2020)\citenamefont
  {Alageshan}, \citenamefont {Verma}, \citenamefont {Bec},\ and\ \citenamefont
  {Pandit}}]{alageshan2020machine}%
  \BibitemOpen
  \bibfield  {author} {\bibinfo {author} {\bibfnamefont {J.~K.}\ \bibnamefont
  {Alageshan}}, \bibinfo {author} {\bibfnamefont {A.~K.}\ \bibnamefont
  {Verma}}, \bibinfo {author} {\bibfnamefont {J.}~\bibnamefont {Bec}}, \ and\
  \bibinfo {author} {\bibfnamefont {R.}~\bibnamefont {Pandit}},\ }\bibfield
  {title} {\enquote {\bibinfo {title} {Machine learning strategies for
  path-planning microswimmers in turbulent flows},}\ }\href@noop {} {\bibfield
  {journal} {\bibinfo  {journal} {Physical Review E}\ }\textbf {\bibinfo
  {volume} {101}},\ \bibinfo {pages} {043110} (\bibinfo {year}
  {2020})}\BibitemShut {NoStop}%
\bibitem [{\citenamefont {Reddy}\ \emph {et~al.}(2016)\citenamefont {Reddy},
  \citenamefont {Celani}, \citenamefont {Sejnowski},\ and\ \citenamefont
  {Vergassola}}]{reddy2016learning}%
  \BibitemOpen
  \bibfield  {author} {\bibinfo {author} {\bibfnamefont {G.}~\bibnamefont
  {Reddy}}, \bibinfo {author} {\bibfnamefont {A.}~\bibnamefont {Celani}},
  \bibinfo {author} {\bibfnamefont {T.~J.}\ \bibnamefont {Sejnowski}}, \ and\
  \bibinfo {author} {\bibfnamefont {M.}~\bibnamefont {Vergassola}},\ }\bibfield
   {title} {\enquote {\bibinfo {title} {Learning to soar in turbulent
  environments},}\ }\href@noop {} {\bibfield  {journal} {\bibinfo  {journal}
  {Proc. Nat. Acad. Sci.}\ }\textbf {\bibinfo {volume} {113}},\ \bibinfo
  {pages} {E4877} (\bibinfo {year} {2016})}\BibitemShut {NoStop}%
\bibitem [{\citenamefont {Colabrese}\ \emph {et~al.}(2017)\citenamefont
  {Colabrese}, \citenamefont {Gustavsson}, \citenamefont {Celani},\ and\
  \citenamefont {Biferale}}]{colabrese2017flow}%
  \BibitemOpen
  \bibfield  {author} {\bibinfo {author} {\bibfnamefont {S.}~\bibnamefont
  {Colabrese}}, \bibinfo {author} {\bibfnamefont {K.}~\bibnamefont
  {Gustavsson}}, \bibinfo {author} {\bibfnamefont {A.}~\bibnamefont {Celani}},
  \ and\ \bibinfo {author} {\bibfnamefont {L.}~\bibnamefont {Biferale}},\
  }\bibfield  {title} {\enquote {\bibinfo {title} {Flow navigation by smart
  microswimmers via reinforcement learning},}\ }\href@noop {} {\bibfield
  {journal} {\bibinfo  {journal} {Phys. Rev. Lett.}\ }\textbf {\bibinfo
  {volume} {118}},\ \bibinfo {pages} {158004} (\bibinfo {year}
  {2017})}\BibitemShut {NoStop}%
\bibitem [{\citenamefont {Verma}\ \emph {et~al.}(2018)\citenamefont {Verma},
  \citenamefont {Novati},\ and\ \citenamefont
  {Koumoutsakos}}]{verma2018efficient}%
  \BibitemOpen
  \bibfield  {author} {\bibinfo {author} {\bibfnamefont {S.}~\bibnamefont
  {Verma}}, \bibinfo {author} {\bibfnamefont {G.}~\bibnamefont {Novati}}, \
  and\ \bibinfo {author} {\bibfnamefont {P.}~\bibnamefont {Koumoutsakos}},\
  }\bibfield  {title} {\enquote {\bibinfo {title} {Efficient collective
  swimming by harnessing vortices through deep reinforcement learning},}\
  }\href@noop {} {\bibfield  {journal} {\bibinfo  {journal} {Proc. Nat. Acad.
  Sci.}\ }\textbf {\bibinfo {volume} {115}},\ \bibinfo {pages} {5849--5854}
  (\bibinfo {year} {2018})}\BibitemShut {NoStop}%
\bibitem [{\citenamefont {Mirzakhanloo}\ \emph {et~al.}(2020)\citenamefont
  {Mirzakhanloo}, \citenamefont {Esmaeilzadeh},\ and\ \citenamefont
  {Alam}}]{mirzakhanloo2020active}%
  \BibitemOpen
  \bibfield  {author} {\bibinfo {author} {\bibfnamefont {M.}~\bibnamefont
  {Mirzakhanloo}}, \bibinfo {author} {\bibfnamefont {S.}~\bibnamefont
  {Esmaeilzadeh}}, \ and\ \bibinfo {author} {\bibfnamefont {M.-R.}\
  \bibnamefont {Alam}},\ }\bibfield  {title} {\enquote {\bibinfo {title}
  {Active cloaking in stokes flows via reinforcement learning},}\ }\href@noop
  {} {\bibfield  {journal} {\bibinfo  {journal} {J. Fluid Mech.}\ }\textbf
  {\bibinfo {volume} {903}} (\bibinfo {year} {2020})}\BibitemShut {NoStop}%
\bibitem [{\citenamefont {Cichos}\ \emph {et~al.}(2020)\citenamefont {Cichos},
  \citenamefont {Gustavsson}, \citenamefont {Mehlig},\ and\ \citenamefont
  {Volpe}}]{cichos2020machine}%
  \BibitemOpen
  \bibfield  {author} {\bibinfo {author} {\bibfnamefont {F.}~\bibnamefont
  {Cichos}}, \bibinfo {author} {\bibfnamefont {K.}~\bibnamefont {Gustavsson}},
  \bibinfo {author} {\bibfnamefont {B.}~\bibnamefont {Mehlig}}, \ and\ \bibinfo
  {author} {\bibfnamefont {G.}~\bibnamefont {Volpe}},\ }\bibfield  {title}
  {\enquote {\bibinfo {title} {Machine learning for active matter},}\
  }\href@noop {} {\bibfield  {journal} {\bibinfo  {journal} {Nature Mach.
  Intel.}\ }\textbf {\bibinfo {volume} {2}},\ \bibinfo {pages} {94} (\bibinfo
  {year} {2020})}\BibitemShut {NoStop}%
\bibitem [{\citenamefont {Qiu}\ \emph {et~al.}(2021)\citenamefont {Qiu},
  \citenamefont {Mousavi}, \citenamefont {Zhao},\ and\ \citenamefont
  {Gustavsson}}]{qiu2021active}%
  \BibitemOpen
  \bibfield  {author} {\bibinfo {author} {\bibfnamefont {J.}~\bibnamefont
  {Qiu}}, \bibinfo {author} {\bibfnamefont {N.}~\bibnamefont {Mousavi}},
  \bibinfo {author} {\bibfnamefont {L.}~\bibnamefont {Zhao}}, \ and\ \bibinfo
  {author} {\bibfnamefont {K.}~\bibnamefont {Gustavsson}},\ }\bibfield  {title}
  {\enquote {\bibinfo {title} {Active gyrotactic stability of microswimmers
  using hydromechanical signals},}\ }\href@noop {} {\bibfield  {journal}
  {\bibinfo  {journal} {Phys. Rev. Fluids}\ } \textbf {\bibinfo {volume} {7}},\ \bibinfo {pages} {014311} (\bibinfo {year}
  {2021})}\BibitemShut {NoStop}%
\bibitem [{\citenamefont {Reddy}\ \emph {et~al.}(2018)\citenamefont {Reddy},
  \citenamefont {Wong-Ng}, \citenamefont {Celani}, \citenamefont {Sejnowski},\
  and\ \citenamefont {Vergassola}}]{reddy2018glider}%
  \BibitemOpen
  \bibfield  {author} {\bibinfo {author} {\bibfnamefont {G.}~\bibnamefont
  {Reddy}}, \bibinfo {author} {\bibfnamefont {J.}~\bibnamefont {Wong-Ng}},
  \bibinfo {author} {\bibfnamefont {A.}~\bibnamefont {Celani}}, \bibinfo
  {author} {\bibfnamefont {T.~J.}\ \bibnamefont {Sejnowski}}, \ and\ \bibinfo
  {author} {\bibfnamefont {M.}~\bibnamefont {Vergassola}},\ }\bibfield  {title}
  {\enquote {\bibinfo {title} {Glider soaring via reinforcement learning in the
  field},}\ }\href@noop {} {\bibfield  {journal} {\bibinfo  {journal} {Nature (Lond.)}\
  }\textbf {\bibinfo {volume} {562}},\ \bibinfo {pages} {236--239} (\bibinfo
  {year} {2018})}\BibitemShut {NoStop}%
\bibitem [{\citenamefont {Mui{\~n}os-Landin}\ \emph {et~al.}(2021)\citenamefont
  {Mui{\~n}os-Landin}, \citenamefont {Fischer}, \citenamefont {Holubec},\ and\
  \citenamefont {Cichos}}]{muinos2021reinforcement}%
  \BibitemOpen
  \bibfield  {author} {\bibinfo {author} {\bibfnamefont {S.}~\bibnamefont
  {Mui{\~n}os-Landin}}, \bibinfo {author} {\bibfnamefont {A.}~\bibnamefont
  {Fischer}}, \bibinfo {author} {\bibfnamefont {V.}~\bibnamefont {Holubec}}, \
  and\ \bibinfo {author} {\bibfnamefont {F.}~\bibnamefont {Cichos}},\
  }\bibfield  {title} {\enquote {\bibinfo {title} {Reinforcement learning with
  artificial microswimmers},}\ }\href@noop {} {\bibfield  {journal} {\bibinfo
  {journal} {Sci. Robot.}\ }\textbf {\bibinfo {volume} {6}} (\bibinfo {year}
  {2021})}\BibitemShut {NoStop}%
\bibitem [{Note1()}]{Note1}%
  \BibitemOpen
  \bibinfo {note} {See supplementary material [url] for supplementary
  figures, details on the implemented Reinforcement Learning algorithm
  including a pseudo-code, and for the captions of supplementary
  movies.}\BibitemShut {Stop}%
\bibitem [{\citenamefont {Lauga}\ and\ \citenamefont
  {Powers}(2009)}]{lauga2009hydrodynamics}%
  \BibitemOpen
  \bibfield  {author} {\bibinfo {author} {\bibfnamefont {E.}~\bibnamefont
  {Lauga}}\ and\ \bibinfo {author} {\bibfnamefont {T.~R.}\ \bibnamefont
  {Powers}},\ }\bibfield  {title} {\enquote {\bibinfo {title} {The
  hydrodynamics of swimming microorganisms},}\ }\href@noop {} {\bibfield
  {journal} {\bibinfo  {journal} {Rep. Progr. Phys.}\ }\textbf {\bibinfo
  {volume} {72}},\ \bibinfo {pages} {096601} (\bibinfo {year}
  {2009})}\BibitemShut {NoStop}%
\bibitem [{\citenamefont {Ishimoto}\ \emph {et~al.}(2020)\citenamefont
  {Ishimoto}, \citenamefont {Gaffney},\ and\ \citenamefont
  {Walker}}]{ishimoto2020regularized}%
  \BibitemOpen
  \bibfield  {author} {\bibinfo {author} {\bibfnamefont {K.}~\bibnamefont
  {Ishimoto}}, \bibinfo {author} {\bibfnamefont {E.~A.}\ \bibnamefont
  {Gaffney}}, \ and\ \bibinfo {author} {\bibfnamefont {B.~J.}\ \bibnamefont
  {Walker}},\ }\bibfield  {title} {\enquote {\bibinfo {title} {Regularized
  representation of bacterial hydrodynamics},}\ }\href@noop {} {\bibfield
  {journal} {\bibinfo  {journal} {Phys. Rev. Fluids}\ }\textbf {\bibinfo
  {volume} {5}},\ \bibinfo {pages} {093101} (\bibinfo {year}
  {2020})}\BibitemShut {NoStop}%
\bibitem [{\citenamefont {Wubbels}\ and\ \citenamefont
  {Schellart}(1997)}]{wubbels1997neuronal}%
  \BibitemOpen
  \bibfield  {author} {\bibinfo {author} {\bibfnamefont {R.~J.}\ \bibnamefont
  {Wubbels}}\ and\ \bibinfo {author} {\bibfnamefont {N.~A.~M.}\ \bibnamefont
  {Schellart}},\ }\bibfield  {title} {\enquote {\bibinfo {title} {Neuronal
  encoding of sound direction in the auditory midbrain of the rainbow trout},}\
  }\href@noop {} {\bibfield  {journal} {\bibinfo  {journal} {J. Neurophysiol.}\
  }\textbf {\bibinfo {volume} {77}},\ \bibinfo {pages} {3060} (\bibinfo {year}
  {1997})}\BibitemShut {NoStop}%
\bibitem [{\citenamefont {Sichert}\ \emph {et~al.}(2009)\citenamefont
  {Sichert}, \citenamefont {Bamler},\ and\ \citenamefont {van
  Hemmen}}]{sichert2009hydrodynamic}%
  \BibitemOpen
  \bibfield  {author} {\bibinfo {author} {\bibfnamefont {A.~B.}\ \bibnamefont
  {Sichert}}, \bibinfo {author} {\bibfnamefont {R.}~\bibnamefont {Bamler}}, \
  and\ \bibinfo {author} {\bibfnamefont {J.~L.}\ \bibnamefont {van Hemmen}},\
  }\bibfield  {title} {\enquote {\bibinfo {title} {Hydrodynamic object
  recognition: When multipoles count},}\ }\href@noop {} {\bibfield  {journal}
  {\bibinfo  {journal} {Phys. Rev. Lett.}\ }\textbf {\bibinfo {volume} {102}},\
  \bibinfo {pages} {058104} (\bibinfo {year} {2009})}\BibitemShut {NoStop}%
\bibitem [{\citenamefont {Jaakkola}\ \emph {et~al.}(1995)\citenamefont
  {Jaakkola}, \citenamefont {Singh},\ and\ \citenamefont
  {Jordan}}]{jaakkola1995reinforcement}%
  \BibitemOpen
  \bibfield  {author} {\bibinfo {author} {\bibfnamefont {T.}~\bibnamefont
  {Jaakkola}}, \bibinfo {author} {\bibfnamefont {S.~P.}\ \bibnamefont {Singh}},
  \ and\ \bibinfo {author} {\bibfnamefont {M.~I.}\ \bibnamefont {Jordan}},\
  }\bibfield  {title} {\enquote {\bibinfo {title} {Reinforcement learning
  algorithm for partially observable markov decision problems},}\ }in\
  \href@noop {} {\emph {\bibinfo {booktitle} {Advances in Neural Information
  Processing Systems}}},\ Vol.~\bibinfo {volume} {8},\ \bibinfo {editor}
  {edited by\ \bibinfo {editor} {\bibfnamefont {D.~S.}\ \bibnamefont
  {Touretzky}}, \bibinfo {editor} {\bibfnamefont {M.~C.}\ \bibnamefont
  {Mozer}}, \ and\ \bibinfo {editor} {\bibfnamefont {M.~E.}\ \bibnamefont
  {Hasselmo}}}\ (\bibinfo  {publisher} {Morgan Kaufmann, San Francisco, CA},\ \bibinfo
  {year} {1995})\ p.\ \bibinfo {pages} {345}\BibitemShut {NoStop}%
\bibitem [{\citenamefont {Bhatnagar}\ \emph {et~al.}(2009)\citenamefont
  {Bhatnagar}, \citenamefont {Sutton}, \citenamefont {Ghavamzadeh},\ and\
  \citenamefont {Lee}}]{bhatnagar2009natural}%
  \BibitemOpen
  \bibfield  {author} {\bibinfo {author} {\bibfnamefont {S.}~\bibnamefont
  {Bhatnagar}}, \bibinfo {author} {\bibfnamefont {R.~S.}\ \bibnamefont
  {Sutton}}, \bibinfo {author} {\bibfnamefont {M.}~\bibnamefont {Ghavamzadeh}},
  \ and\ \bibinfo {author} {\bibfnamefont {M.}~\bibnamefont {Lee}},\ }\bibfield
   {title} {\enquote {\bibinfo {title} {Natural actor--critic algorithms},}\
  }\href@noop {} {\bibfield  {journal} {\bibinfo  {journal} {Automatica}\
  }\textbf {\bibinfo {volume} {45}},\ \bibinfo {pages} {2471--2482} (\bibinfo
  {year} {2009})}\BibitemShut {NoStop}%
\bibitem [{\citenamefont {Grondman}\ \emph {et~al.}(2012)\citenamefont
  {Grondman}, \citenamefont {Busoniu}, \citenamefont {Lopes},\ and\
  \citenamefont {Babuska}}]{grondman2012survey}%
  \BibitemOpen
  \bibfield  {author} {\bibinfo {author} {\bibfnamefont {I.}~\bibnamefont
  {Grondman}}, \bibinfo {author} {\bibfnamefont {L.}~\bibnamefont {Busoniu}},
  \bibinfo {author} {\bibfnamefont {G.~A.~D.}\ \bibnamefont {Lopes}}, \ and\
  \bibinfo {author} {\bibfnamefont {R.}~\bibnamefont {Babuska}},\ }\bibfield
  {title} {\enquote {\bibinfo {title} {A survey of actor-critic reinforcement
  learning: Standard and natural policy gradients},}\ }\href@noop {} {\bibfield
   {journal} {\bibinfo  {journal} {IEEE Trans. Syst. Man. Cybern. Part C}\
  }\textbf {\bibinfo {volume} {42}},\ \bibinfo {pages} {1291--1307} (\bibinfo
  {year} {2012})}\BibitemShut {NoStop}%
\bibitem [{\citenamefont {Hennes}\ \emph {et~al.}(2019)\citenamefont {Hennes},
  \citenamefont {Morrill}, \citenamefont {Omidshafiei}, \citenamefont {Munos},
  \citenamefont {Perolat}, \citenamefont {Lanctot}, \citenamefont {Gruslys},
  \citenamefont {Lespiau}, \citenamefont {Parmas}, \citenamefont
  {Duenez-Guzman},\ and\ \citenamefont {Tuyls}}]{hennes2019neural}%
  \BibitemOpen
  \bibfield  {author} {\bibinfo {author} {\bibfnamefont {D.}~\bibnamefont
  {Hennes}}, \bibinfo {author} {\bibfnamefont {D.}~\bibnamefont {Morrill}},
  \bibinfo {author} {\bibfnamefont {S.}~\bibnamefont {Omidshafiei}}, \bibinfo
  {author} {\bibfnamefont {R.}~\bibnamefont {Munos}}, \bibinfo {author}
  {\bibfnamefont {J.}~\bibnamefont {Perolat}}, \bibinfo {author} {\bibfnamefont
  {M.}~\bibnamefont {Lanctot}}, \bibinfo {author} {\bibfnamefont
  {A.}~\bibnamefont {Gruslys}}, \bibinfo {author} {\bibfnamefont {J.-B.}\
  \bibnamefont {Lespiau}}, \bibinfo {author} {\bibfnamefont {P.}~\bibnamefont
  {Parmas}}, \bibinfo {author} {\bibfnamefont {E.}~\bibnamefont
  {Duenez-Guzman}}, \ and\ \bibinfo {author} {\bibfnamefont {K.}~\bibnamefont
  {Tuyls}},\ }\bibfield  {title} {\enquote {\bibinfo {title} {Neural replicator
  dynamics},}\ }\href@noop {} {\bibfield  {journal} {\bibinfo  {journal}
  {arXiv:1906.00190 [cs.LG]}\ } (\bibinfo {year} {2019})}\BibitemShut {NoStop}%
\bibitem [{\citenamefont {Amari}(1998)}]{amari1998natural}%
  \BibitemOpen
  \bibfield  {author} {\bibinfo {author} {\bibfnamefont {S.-I.}\ \bibnamefont
  {Amari}},\ }\bibfield  {title} {\enquote {\bibinfo {title} {Natural gradient
  works efficiently in learning},}\ }\href@noop {} {\bibfield  {journal}
  {\bibinfo  {journal} {Neural Comput.}\ }\textbf {\bibinfo {volume} {10}},\
  \bibinfo {pages} {251--276} (\bibinfo {year} {1998})}\BibitemShut {NoStop}%
\bibitem [{\citenamefont {Eaton}\ \emph {et~al.}(2001)\citenamefont {Eaton},
  \citenamefont {Lee},\ and\ \citenamefont {Foreman}}]{eaton2001mauthner}%
  \BibitemOpen
  \bibfield  {author} {\bibinfo {author} {\bibfnamefont {R.~C.}\ \bibnamefont
  {Eaton}}, \bibinfo {author} {\bibfnamefont {R.~K.~K.}\ \bibnamefont {Lee}}, \
  and\ \bibinfo {author} {\bibfnamefont {M.~B.}\ \bibnamefont {Foreman}},\
  }\bibfield  {title} {\enquote {\bibinfo {title} {The {M}authner cell and
  other identified neurons of the brainstem escape network of fish},}\
  }\href@noop {} {\bibfield  {journal} {\bibinfo  {journal} {Progr.
  Neurobiol.}\ }\textbf {\bibinfo {volume} {63}},\ \bibinfo {pages} {467--485}
  (\bibinfo {year} {2001})}\BibitemShut {NoStop}%
\bibitem [{\citenamefont {Arulkumaran}\ \emph {et~al.}(2017)\citenamefont
  {Arulkumaran}, \citenamefont {Deisenroth}, \citenamefont {Brundage},\ and\
  \citenamefont {Bharath}}]{arulkumaran2017deep}%
  \BibitemOpen
  \bibfield  {author} {\bibinfo {author} {\bibfnamefont {K.}~\bibnamefont
  {Arulkumaran}}, \bibinfo {author} {\bibfnamefont {M.~P.}\ \bibnamefont
  {Deisenroth}}, \bibinfo {author} {\bibfnamefont {M.}~\bibnamefont
  {Brundage}}, \ and\ \bibinfo {author} {\bibfnamefont {A.~A.}\ \bibnamefont
  {Bharath}},\ }\bibfield  {title} {\enquote {\bibinfo {title} {Deep
  reinforcement learning: A brief survey},}\ }\href@noop {} {\bibfield
  {journal} {\bibinfo  {journal} {IEEE Signal Proces. Mag.}\ }\textbf {\bibinfo
  {volume} {34}},\ \bibinfo {pages} {26--38} (\bibinfo {year}
  {2017})}\BibitemShut {NoStop}%
\bibitem [{Note2()}]{Note2}%
  \BibitemOpen
  \bibinfo {note} {Note that each point represents the average over the
  previous and following 50 episodes, so it is not immediate to recognize those
  episodes in which the evader wins, i.e., in which $T/T_{max}=1$.}\BibitemShut
  {Stop}%
\bibitem [{Note3()}]{Note3}%
  \BibitemOpen
  \bibinfo {note} {We use a fixed learning rate instead of an adaptive one, and
  possibly, due to the need to explore, a larger number of episodes per turn
  would be necessary.}\BibitemShut {Stop}%
\bibitem [{\citenamefont {Belkhouche}\ \emph {et~al.}(2007)\citenamefont
  {Belkhouche}, \citenamefont {Belkhouche},\ and\ \citenamefont
  {Rastgoufard}}]{belkhouche2007parallel}%
  \BibitemOpen
  \bibfield  {author} {\bibinfo {author} {\bibfnamefont {F.}~\bibnamefont
  {Belkhouche}}, \bibinfo {author} {\bibfnamefont {B.}~\bibnamefont
  {Belkhouche}}, \ and\ \bibinfo {author} {\bibfnamefont {P.}~\bibnamefont
  {Rastgoufard}},\ }\bibfield  {title} {\enquote {\bibinfo {title} {Parallel
  navigation for reaching a moving goal by a mobile robot},}\ }\href@noop {}
  {\bibfield  {journal} {\bibinfo  {journal} {Robotica}\ }\textbf {\bibinfo
  {volume} {25}},\ \bibinfo {pages} {63--74} (\bibinfo {year}
  {2007})}\BibitemShut {NoStop}%
\bibitem [{\citenamefont {Adam}\ and\ \citenamefont
  {Delbr{\"u}ck}(1968)}]{adam1968reduction}%
  \BibitemOpen
  \bibfield  {author} {\bibinfo {author} {\bibfnamefont {G}~\bibnamefont
  {Adam}}\ and\ \bibinfo {author} {\bibfnamefont {M}~\bibnamefont
  {Delbr{\"u}ck}},\ }\bibfield  {title} {\enquote {\bibinfo {title} {Reduction
  of dimensionality in biological diffusion processes},}\ }\href@noop {}
  {\bibfield  {journal} {\bibinfo  {journal} {Structural chemistry and
  molecular biology}\ }\textbf {\bibinfo {volume} {198}},\ \bibinfo {pages}
  {198--215} (\bibinfo {year} {1968})}\BibitemShut {NoStop}%
\bibitem [{\citenamefont {Olberg}\ \emph {et~al.}(2000)\citenamefont {Olberg},
  \citenamefont {Worthington},\ and\ \citenamefont {Venator}}]{olberg2000prey}%
  \BibitemOpen
  \bibfield  {author} {\bibinfo {author} {\bibfnamefont {R.~M.}\ \bibnamefont
  {Olberg}}, \bibinfo {author} {\bibfnamefont {A.~H.}\ \bibnamefont
  {Worthington}}, \ and\ \bibinfo {author} {\bibfnamefont {K.~R.}\ \bibnamefont
  {Venator}},\ }\bibfield  {title} {\enquote {\bibinfo {title} {Prey pursuit
  and interception in dragonflies},}\ }\href@noop {} {\bibfield  {journal}
  {\bibinfo  {journal} {J. Compar. Physiol. A}\ }\textbf {\bibinfo {volume}
  {186}},\ \bibinfo {pages} {155} (\bibinfo {year} {2000})}\BibitemShut
  {NoStop}%
\bibitem [{\citenamefont {Chiu}\ \emph {et~al.}(2010)\citenamefont {Chiu},
  \citenamefont {Reddy}, \citenamefont {Xian}, \citenamefont {Krishnaprasad},\
  and\ \citenamefont {Moss}}]{chiu2010effects}%
  \BibitemOpen
  \bibfield  {author} {\bibinfo {author} {\bibfnamefont {C.}~\bibnamefont
  {Chiu}}, \bibinfo {author} {\bibfnamefont {P.~V.}\ \bibnamefont {Reddy}},
  \bibinfo {author} {\bibfnamefont {W.}~\bibnamefont {Xian}}, \bibinfo {author}
  {\bibfnamefont {P.~S.}\ \bibnamefont {Krishnaprasad}}, \ and\ \bibinfo
  {author} {\bibfnamefont {C.~F.}\ \bibnamefont {Moss}},\ }\bibfield  {title}
  {\enquote {\bibinfo {title} {Effects of competitive prey capture on flight
  behavior and sonar beam pattern in paired big brown bats, eptesicus
  fuscus},}\ }\href@noop {} {\bibfield  {journal} {\bibinfo  {journal} {J.
  Exper. Biol.}\ }\textbf {\bibinfo {volume} {213}},\ \bibinfo {pages} {3348}
  (\bibinfo {year} {2010})}\BibitemShut {NoStop}%
\bibitem [{\citenamefont {Lanchester}\ and\ \citenamefont
  {Mark}(1975)}]{lanchester1975pursuit}%
  \BibitemOpen
  \bibfield  {author} {\bibinfo {author} {\bibfnamefont {B.~S.}\ \bibnamefont
  {Lanchester}}\ and\ \bibinfo {author} {\bibfnamefont {R.~F.}\ \bibnamefont
  {Mark}},\ }\bibfield  {title} {\enquote {\bibinfo {title} {Pursuit and
  prediction in the tracking of moving food by a teleost fish (acanthaluteres
  spilomelanurus)},}\ }\href@noop {} {\bibfield  {journal} {\bibinfo  {journal}
  {J. Exper. Biol.}\ }\textbf {\bibinfo {volume} {63}},\ \bibinfo {pages} {627}
  (\bibinfo {year} {1975})}\BibitemShut {NoStop}%
\bibitem [{\citenamefont {Kottapalli}\ \emph {et~al.}(2015)\citenamefont
  {Kottapalli}, \citenamefont {Asadnia}, \citenamefont {Miao},\ and\
  \citenamefont {Triantafyllou}}]{kottapalli2015soft}%
  \BibitemOpen
  \bibfield  {author} {\bibinfo {author} {\bibfnamefont {A.~G.~P.}\
  \bibnamefont {Kottapalli}}, \bibinfo {author} {\bibfnamefont
  {M.}~\bibnamefont {Asadnia}}, \bibinfo {author} {\bibfnamefont
  {J.}~\bibnamefont {Miao}}, \ and\ \bibinfo {author} {\bibfnamefont
  {M.}~\bibnamefont {Triantafyllou}},\ }\bibfield  {title} {\enquote {\bibinfo
  {title} {Soft polymer membrane micro-sensor arrays inspired by the
  mechanosensory lateral line on the blind cavefish},}\ }\href@noop {}
  {\bibfield  {journal} {\bibinfo  {journal} {J. Intell. Mat. Syst. Struct.}\
  }\textbf {\bibinfo {volume} {26}},\ \bibinfo {pages} {38} (\bibinfo {year}
  {2015})}\BibitemShut {NoStop}%
\bibitem [{\citenamefont {Free}\ \emph {et~al.}(2020)\citenamefont {Free},
  \citenamefont {Lee},\ and\ \citenamefont {Paley}}]{free2020bioinspired}%
  \BibitemOpen
  \bibfield  {author} {\bibinfo {author} {\bibfnamefont {B.~A.}\ \bibnamefont
  {Free}}, \bibinfo {author} {\bibfnamefont {J.}~\bibnamefont {Lee}}, \ and\
  \bibinfo {author} {\bibfnamefont {D.~A.}\ \bibnamefont {Paley}},\ }\bibfield
  {title} {\enquote {\bibinfo {title} {Bioinspired pursuit with a swimming
  robot using feedback control of an internal rotor},}\ }\href@noop {}
  {\bibfield  {journal} {\bibinfo  {journal} {Bioinsp. Biomim.}\ }\textbf
  {\bibinfo {volume} {15}},\ \bibinfo {pages} {035005} (\bibinfo {year}
  {2020})}\BibitemShut {NoStop}%
\bibitem [{\citenamefont {Marchese}\ \emph {et~al.}(2014)\citenamefont
  {Marchese}, \citenamefont {Onal},\ and\ \citenamefont
  {Rus}}]{marchese2014autonomous}%
  \BibitemOpen
  \bibfield  {author} {\bibinfo {author} {\bibfnamefont {A.~D.}\ \bibnamefont
  {Marchese}}, \bibinfo {author} {\bibfnamefont {C.~D.}\ \bibnamefont {Onal}},
  \ and\ \bibinfo {author} {\bibfnamefont {D.}~\bibnamefont {Rus}},\ }\bibfield
   {title} {\enquote {\bibinfo {title} {Autonomous soft robotic fish capable of
  escape maneuvers using fluidic elastomer actuators},}\ }\href@noop {}
  {\bibfield  {journal} {\bibinfo  {journal} {Soft Robot.}\ }\textbf {\bibinfo
  {volume} {1}},\ \bibinfo {pages} {75} (\bibinfo {year} {2014})}\BibitemShut
  {NoStop}%
\end{thebibliography}
%

\end{document}